\documentclass[useAMS,usenatbib]{mn2e}

\usepackage{epsfig, amsmath, natbib, setspace}

\title[Tidal Evolution, Evaporative Mass Loss and CoRoT-7 b]
{The Roles of Tidal Evolution and Evaporative Mass Loss in the Origin of CoRoT-7 b}
\author[B. Jackson et al.]{Brian Jackson$^{1,2}$\thanks{E-mail: brian.k.jackson@nasa.gov}, Neil Miller$^{3}$, Rory Barnes$^{4}$, Sean N. Raymond$^{5, 6}$, \newauthor Jonathan Fortney$^{3}$, and Richard Greenberg$^{7}$\\
$^{1}$NASA Goddard Space Flight Center, Greenbelt MD 20771\\
$^{2}$NASA Postdoctoral Program Fellow\\
$^{3}$Department of Astronomy and Astrophysics, University of California, Santa Cruz, CA 95064\\
$^{4}$Department of Astronomy, University of Washington, Seattle, WA, 98195-1580\\
$^{5}$Universit\'{e} de Bordeaux, Observatoire Aquitain des Sciences de l'Univers, 2 rue de l'Observatoire, BP 89, F-33271 Floirac Cedex, France\\
$^{6}$CNRS, UMR 5804, Laboratoire d'Astrophysique de Bordeaux, 2 rue de l'Observatoire, BP 89, F-33271 Floirac Cedex, France\\
$^{7}$Lunar and Planetary Laboratory, University of Arizona, Tucson AZ 85721}
\date{Accepted 2010 May 7.  Received 2010 April 16; in original form 2010 February 21}

\pagerange{\pageref{firstpage}--\pageref{lastpage}} \pubyear{2010}

\begin{document}

\label{firstpage}

\maketitle

\begin{abstract}
CoRoT-7 b is the first confirmed rocky exoplanet, but, with an orbital semi-major axis of 0.0172 AU, its origins may be unlike any rocky planet in our solar system. In this study, we consider the roles of tidal evolution and evaporative mass loss in CoRoT-7 b's history, which together have modified the planet's mass and orbit. If CoRoT-7 b has always been a rocky body, evaporation may have driven off almost half its original mass, but the mass loss may depend sensitively on the extent of tidal decay of its orbit. As tides caused CoRoT-7 b's orbit to decay, they brought the planet closer to its host star, thereby enhancing the mass loss rate. Such a large mass loss also suggests the possibility that CoRoT-7 b began as a gas giant planet and had its original atmosphere completely evaporated. In this case, we find that CoRoT-7 b's original mass probably didn't exceed 200 Earth masses (about 2/3 of a Jupiter mass). Tides raised on the host star by the planet may have significantly reduced the orbital semi-major axis, perhaps causing the planet to migrate through mean-motion resonances with the other planet in the system, CoRoT-7 c. The coupling between tidal evolution and mass loss may be important not only for CoRoT-7 b but also for other close-in exoplanets, and future studies of mass loss and orbital evolution may provide insight into the origin and fate of close-in planets, both rocky and gaseous.
\end{abstract}


\section{Introduction}
\indent The recently discovered exoplanet CoRoT-7 b has a mass of $M_p = 4.8 \pm 0.8$ Earth masses ($M_{Earth}$) and a radius of only $R_p = 1.68 \pm 0.09$ Earth radii ($R_{Earth}$). Thus its density likely lies between 0.71 and 1.4 times Earth's density, which suggests CoRoT-7 b is made of similar materials \citep{2009A&A...506..287L, 2009A&A...506..303Q}. It appears to be the first example of a rocky exoplanet. However, with a semi-major axis $a$ of only $0.0172 \pm 0.00029$ AU and $e \approx 0$ \citep{2009A&A...506..287L}, CoRoT-7 b has an orbit unlike the rocky planets in our solar system. CoRoT-7 b's surface may even be partially melted as a result of the strong stellar insolation \citep{2009A&A...506..287L}. (Consequently, when we refer to CoRoT-7 b's ``rocky'' surface, we are referring to the composition and do not mean to imply that the surface is solid rock.) The announcement of an additional, nearby planet in the system, CoRoT-7 c, ($M_p = 8.4 \pm 0.9$ $M_{Earth}$, $a = 0.046$ AU, and $e \approx 0$) \citep{2009A&A...506..303Q} also suggests that the system has an interesting dynamical history. \\
\indent Missions like \emph{Kepler} \citep{2008IAUS..249...17B} and \emph{CoRoT} \citep{2006cosp...36.3749B} may find dozens of additional rocky exoplanets through transit observations. Although Kepler is expected to find several rocky planets in orbits like the Earth's \citep{2008IAUS..249...17B}, the first and most easily detectable rocky planets will have orbits closer to their stars \citep{2009PASP..121..952D}. Thus a detailed understanding of such planets will be critical to understanding the planets first discovered by these space-based missions. \\
\indent The origins of these planets may be very different from the origins of rocky planets in our solar system. It may be difficult for rocky exoplanets to form as close to their stars as CoRoT-7 b and c because high temperatures in the protoplanetary gas disk may inhibit condensation and accretion of solid materials (\emph{e.g.},  \citealt{1994ApJ...421..615P}). At the very least, solid planets that form close-in will probably be depleted in volatile materials (\emph{e.g.}, water [\citealt{2008MNRAS.384..663R}]). Instead of forming in their current orbits, close-in rocky exoplanets like CoRoT-7 b may have coalesced in orbits farther away from their host stars, where disk temperatures were lower, and were brought to their current orbits. A variety of processes for bringing planets to such close-in orbits have been proposed \citep{2004ApJ...614..955M, 2007Sci...318..210G, 2008MNRAS.384..663R}, and the orbital architecture of systems with close-in planets and the planets' physical properties provide clues to their histories \citep{2008MNRAS.384..663R}. \\
\indent One process that may have been important during CoRoT-7 b's history is evaporative mass loss. Evaporation is thought to influence gaseous exoplanets made primarily of H and He. For example, observations of HD 209458 b suggest the planet's atmosphere is evaporating at a rate of order 0.1 $M_{Earth}$/Gyr \citep{2003Natur.422..143V,2004ApJ...604L..69V}, in line with theoretical expectations from studies of evaporative mass loss \citep{2003ApJ...598L.121L, 2004A&A...418L...1L, 2004Icar..170..167Y, 2005ApJ...621.1049T, 2008SSRv..139..437Y}. Moreover, gaps in the distributions of orbital and physical properties of close-in planets have also been interpreted to indicate that many gaseous planets have had their atmospheres completely evaporated \citep{2009MNRAS.396.1012D}. The complete removal of a gas giant's atmosphere would likely leave behind its rocky core, with a mass of perhaps several $M_{Earth}$ \citep{2004A&A...419L..13B, 2008MNRAS.384..663R}. These considerations suggest there may be a large population of close-in planetary bodies that are remnant cores of evaporated gas giants \citep{2005A&A...436L..47B, 2008PhDT........15H}. On the other hand, other studies argue that the observations of HD 209458 b do not imply significant loss of mass \citep{2007ApJ...671L..61B, 2008Natur.451..970H}, and a theoretical study by \citet{2009ApJ...693...23M} suggested that complete evaporation of a gas giant's atmosphere is unlikely. These competing hypotheses may now be testable. With the detection capabilities of the \emph{Kepler} and \emph{CoRoT} missions, rocky planets arising from a variety of histories may be detected. \\
\indent Evaporation of mass may continue even after the planet loses its original H/He envelope. \citet{2009arXiv0907.3067V} suggest the strong insolation received by CoRoT-7 b may be sufficient to have evaporated and removed several Earth masses of material, if the planet has always been solid (made of ice and/or rock). \citet{2009ApJ...703L.113S} showed that thermal evaporation of rocky material from CoRoT-7 b's surface may produce a tenuous exosphere, similar to Mercury's. In this study, we consider evaporation both of a putative gaseous envelope and of solid material during CoRoT-7 b's history, including effects of its evolving orbit.\\
\indent In addition to mass loss, tides have played an important and interrelated role in CoRoT-7 b's history. Tides have shaped the distribution of orbital elements for close-in exoplanets, reducing eccentricities \citep{1996ApJ...470.1187R, 2006ApJ...638L..45F, 2008ApJ...678.1396J} and semi-major axes \citep{2009MNRAS.395.2268B, 2009ApJ...698.1357J, 2009ApJ...692L...9L}. Tides may also help to bring rocky planets into close-in orbits \citep{2008MNRAS.384..663R}. In fact, going forward in time, CoRoT-7 b may migrate into its host star in less than a few billion years, and it may have undergone past orbital migration \citep{2009ApJ...698.1357J}. Also, CoRoT-7 b likely has undergone or currently undergoes strong tidal heating, either because its original orbit was eccentric or interactions with nearby CoRoT-7 c keep the eccentricity non-zero \citep{2010ApJ...709L..95B}. Tidal evolution of its orbit may have been even faster in the past if CoRoT-7 b once had a massive gaseous envelope because the rate of evolution of a circular orbit scales with its mass. As the planet lost mass, though, the tidal evolution may have slowed as the planet reached its current orbit.\\
\indent In this paper, we study the coupling of evaporative mass loss and tidal evolution on CoRoT-7 b. Using parameterized models for mass loss and orbital decay due to tides, we evolved CoRoT-7 b's mass and orbit backward (and forward) in time to determine the range of original masses and orbits consistent with the planet's current mass and orbit. Although the exact initial conditions depend sensitively on the chosen model parameters, we considered a wide range of parameters to map out all the possibilities. We find that, if CoRoT-7 b has always been a solid planet, it may have lost as much as half its original mass and had its semi-major axis reduced by as much as 60\% by tidal migration alone. If CoRoT-7 b was originally a gas giant planet and only recently lost its gaseous envelope, its original mass did not exceed 200 $M_{Earth}$, and its original semi-major axis could have been almost twice as large as the current value. In both scenarios, CoRoT-7 b may have encountered and migrated through a 3:1 mean motion resonance with nearby CoRoT-7 c, as tides caused planet b's orbit to decay. Consideration of the resulting dynamical interactions may place constraints on the orbital history of the CoRoT-7 system and the original mass of CoRoT-7 b. We also discuss how the coupling between orbital decay and mass loss may play an important in the evolution and survival of gaseous close-in planets.
\section{Models}
\indent Several recent studies of evaporation of exoplanetary atmospheres provide simple, parameterized models for mass loss rates (\emph{e.g.}, \citealt{2004A&A...419L..13B,2005A&A...439..771J, 2007A&A...472..329E}). Other authors have applied conventional models for tidal damping to study the orbital evolution of close-in exoplanets \citep{1963MNRAS.126..257G, 1966Icar....5..375G, 1996ApJ...470.1187R, 2008CeMDA.101..171F, 2008ApJ...678.1396J}. Although the two processes may be coupled, few studies have considered both processes together. \citet{2003ApJ...588..509G} investigated the coupling between tidal heating and Roche lobe overflow, but they only considered the orbital effects of the tide raised on the planet by the host star. For close-in exoplanets on nearly circular orbits, like CoRoT-7 b, the tide raised on the host star by the planet can dominate the orbital evolution. In this section, we describe our model for coupled mass-orbital evolution and how we can use it to constrain the original orbit and mass of CoRoT-7 b.\\
\subsection{Evaporative Mass Loss Model}
\begin{figure}
\includegraphics[width=8in,angle=-90]{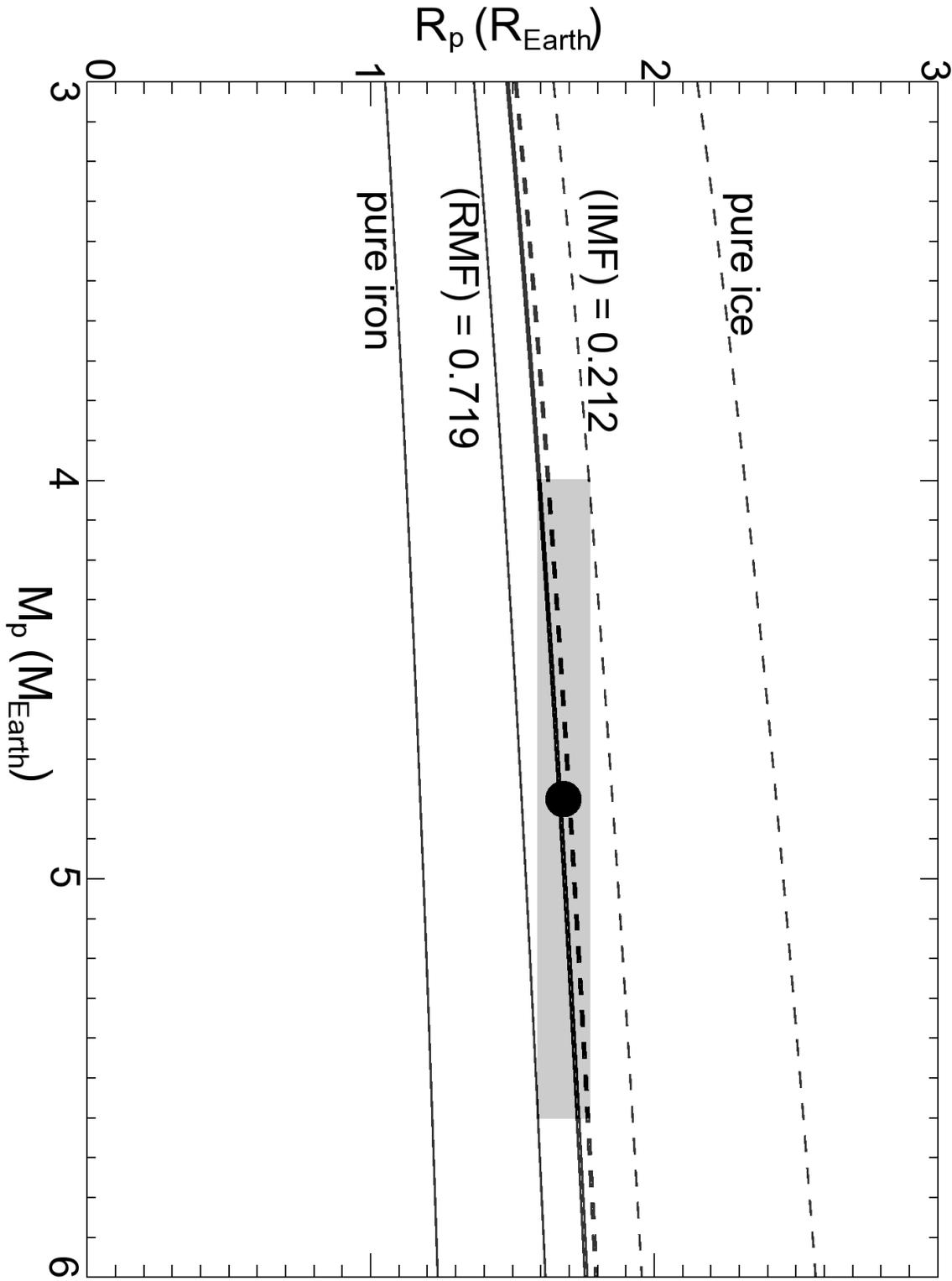}
\caption{Range of planetary compositions consistent with CoRoT-7 b's mass $M_p$ and radius $R_p$, as reported in \citet{2009A&A...506..287L}. The filled circle represents the nominal $M_p$ and $R_p$, while the shaded rectangle shows the uncertainties. Solid lines represent the $M_p$-$R_p$ relation if the planet is made of rock and/or iron, while the dashed represent the relation if the planet is made of rock and/or ice. From bottom to top, the solid lines represent rock mass fractions (RMF) of 0 (pure iron), 0.719, and 0.983. From top to bottom, the dashed lines represent ice mass fractions (IMF) of 1, 0.212, and 0.0108.}
\label{fig:plot_Rp_v_Mp_mf}
\end{figure}

\indent For energy-limited evaporative mass loss, the rate at which escaping gas molecules carry away energy from the planet is roughly proportional to the rate of input of energy from stellar insolation \citep{2008SSRv..139..437Y}. Relating the rate of energy input from insolation to the change in gravitational energy required for a gas molecule to escape yields estimates of mass loss rates. We take the mass loss rate to be \citep{2007A&A...472..329E}:
\begin{equation}
\frac{dM_p}{dt} = -\frac{\pi R_p^3 \epsilon F_{xuv}}{G M_p K_{tide}}
\label{eqn:dMpdt}
\end{equation}
where $F_{xuv}$ is the extreme UV (XUV) flux from the star (wavelengths from 0.1 to 100 nm), evaluated at the planet's orbital distance. $R_p$ is the planet's radius, and G is the gravitational constant. This equation assumes that the planet's optical cross-section in the XUV is $\pi R_p^2$, as suggested by \citet{2007A&A...472..329E} and \citet{2004Icar..170..167Y}. The factors $\epsilon$ and $K_{tide}$ (each of which may range from 0 to 1) are discussed next.\\
\indent The parameter $\epsilon$ represents the fraction of the incoming energy that is carried away by the escaping gas. Some of the incoming energy may drive chemistry or be lost through radiative cooling of the planet, rather than driving escape. In these cases, $\epsilon$ is less than 1 \citep{2007ApJ...658L..59H, 2007Icar..187..358H, 2004Icar..170..167Y, 2009A&A...506..399L}. For example, estimates of the mass loss rate for HD 209458 b suggest $\frac{dM_p}{dt} \sim 4\times10^{10}$ g/s (\emph{e.g.}, \citealt{2008SSRv..139..437Y}), which corresponds to $\epsilon \sim 0.4$. On the other hand, an analysis of atmospheric escape from Venus suggests $0.15$ may be a more appropriate value for $\epsilon$ \citep{1996JGR...10126039C}. \citet{2009A&A...506..399L} indicate that $\epsilon$ is probably less than 0.6 for hot Jupiters but will depend on the host star's activity and the planet's atmospheric composition, among other things. Moreover, vaporization and removal of Mercury's exosphere involve processes that may be different from evaporation of gas giant planets \citep{2007SSRv..132..433K}, and evolution of the evaporating planet's composition may result in a time-dependent $\epsilon$-value \citep{2009ApJ...693...23M}. However, Equation \ref{eqn:dMpdt} provides a reasonable estimate of evaporation of planetary mass by XUV flux, which plays an important, if not dominant, role for Mercury as well as gaseous planets. Here, we consider a range of time-invariant $\epsilon$ values: 0, 0.1, 0.25, 0.5, and 1. We include $\epsilon = 0$ to compare orbital evolution rates with and without mass loss. Note that we do not include any possible dependence of $\epsilon$ on the solid planet's composition, even though mass loss may be more rapid if the planet were more enriched in volatiles. The detailed modeling required to quantify that effect is beyond the scope of this study. \\
\indent For planets close to their host stars, gas molecules may escape from a planet even with a velocity significantly less than the usual escape velocity because they only have to reach the Roche lobe. This reduction in the required velocity results in an increased escape rate for a given rate of energy input. In Equation \ref{eqn:dMpdt}, $K_{tide}$ represents this reduction in required escape energy \citep{2007A&A...472..329E}. $K_{tide}$ is given by:
\begin{equation}
K_{tide} = 1 - \frac{3}{2\xi} + \frac{1}{2\xi^3}
\label{eqn:Ktide}
\end{equation}
where $\xi$ is the ratio of the Roche radius ($=(M_p/3M_*)^{1/3} a$) to $R_p$. For a planet far from its host star, where the Roche radius is much greater than $R_p$, $K_{tide}$ approaches one, indicating no enhancement of the escape rate. For HD 209458 b, with $a = 0.047$ AU, $K_{tide} = 0.65$. As tides bring a planet closer and closer to its host star, $K_{tide}$ decreases to reflect the reduction in required escape velocity.\\
\indent In employing this formulation for the mass loss rate, we treat its effects as if the mass loss is isotropic, that is, as if the flux of escaping gas is the same in all directions. In reality, gas may escape preferentially through the L1 Lagrange point between the planet and star, where the net acceleration is close to zero \citep{2003ApJ...588..509G}, which may affect the planet's orbital evolution. However, incorporating this directionality may require a 3-D hydrodynamic model and so is beyond the scope of our study. We discuss this point more in Section 4.\\
\indent We also include the evolution of the stellar XUV flux in our model. The stellar XUV flux powers mass loss, but as stars age, typically their rotation rates drop with time, which reduces their magnetic activity and XUV radiation (\emph{e.g.}, \citealt{1972ApJ...171..565S}). \citet{2005ApJ...622..680R} studied the evolution of the XUV flux from G-type stars like CoRoT-7 and suggested $F_{xuv}$ evolves according to: 
\begin{equation}
F_{xuv} = \lambda t_{Gyr}^{-\beta}/a^2
\label{eqn:Fxuv}
\end{equation}
where $\lambda = 29.7$ ergs s$^{-1}$ cm$^{-2}$, $\beta = 1.23$, $t_{Gyr}$ is the star's age in billions of years and $a$ is in AU. This equation is an empirical fit to data for observations of stars with ages greater than $0.1$ Gyr and therefore may only apply for $t > 0.1$ Gyr. However, planets in our solar system (and, presumably, in others) developed atmospheres in less than a few tens of millions of years after formation \citep{1986JGR....91..291A}. Also, before $t = 0.1$ Gyr, CoRoT-7's XUV flux was probably somewhat larger than at later times. Thus, atmospheric loss may also have been significant at earlier times than considered here. In this sense, our results provide a lower limit to the total mass lost by CoRoT-7 b. However, that earlier period of time was much shorter than the time span we consider, and so the total mass lost may have been not much greater than suggested by our calculations. Without more data on stellar XUV at early times, stopping our models at $t = 0.1$ Gyr is probably an adequate approximation for our purposes.\\
\indent CoRoT-7 b receives sufficient insolation that vaporization of its rocky surface may produce a tenuous atmosphere (\emph{e.g.}, \citealt{2009ApJ...703L.113S}). \citet{2009arXiv0907.3067V} suggested that the rate of vaporization of rocky material may be orders of magnitude larger than the atmospheric mass loss rate, sufficiently rapid to continually resupply the atmosphere, so in our model, we allow evaporation to continue even after a model planet loses its entire original H/He envelope.\\
\indent To determine the planet's radius after the original H/He envelope is lost, we use Equations 7 \& 8 from \citet{2007ApJ...659.1661F}, along with corrections given in \citet{2007ApJ...668.1267F}. These equations provide mass-radius relations for solid planets made of ice and rock or rock and iron, which requires us to make some assumption about the planet's composition in order to determine its radius, given a mass. \\
\indent Figure \ref{fig:plot_Rp_v_Mp_mf} shows the range of compositions allowed by CoRoT-7 b's observed mass and radius based on the equations from \citet{2007ApJ...659.1661F}. That figure shows that, if made of ice and rock, CoRoT-7 b's ice mass fraction (IMF) may lie between 0.01 and 0.212. If made of rock and iron, CoRoT-7 b's rock mass fraction (RMF) may lie in the range 0.719 to 0.983. These latter numbers are similar to compositions for CoRoT-7 b proposed by \citet{2009arXiv0907.3067V}. In our models, we consider the allowed minimum, middle and maximum values for both the IMF and RMF. However, the results are much less sensitive to our choice of IMF or RMF than to the other model parameters. \\
\indent Given the uncertainty in current composition and structure of CoRoT-7 b, our model of coupled orbit evolution and evaporative mass loss cannot uniquely determine the mass of the gaseous envelope that CoRoT-7 b may originally have had. CoRoT-7 b may have begun as a gas giant with a core mass slightly larger than CoRoT-7 b's current mass (say, 7 $M_{Earth}$). After evaporation stripped the original atmosphere, it could also remove $\sim$ 1 $M_{Earth}$ from the original core to give the mass seen today. Consequently, in order to test the hypothesis that CoRoT-7 b started out as a gas giant, we consider a range of masses for the core of the original planet ($M_{core}$), ranging from CoRoT-7 b's current mass (in which case, the planet very recently lost its original atmosphere) up to 19 $M_{Earth}$. In our simulations, we run time backward, allowing the solid mass of the planet to increase until $M_p$ reaches our assumed $M_{core}$. From that point continuing back in time, we assume the planet ``gains'' mass in the form of H/He gas.\\
\indent Once the planet has a thick enough atmosphere, we define the planet's radius $R_p$ to be the altitude level at which most of the stellar XUV is absorbed \citep{2007A&A...472..329E, 2004Icar..170..167Y, 2009A&A...506..399L}. For $M_p \ge 0.018 M_{Jup}$, $R_p$ corresponds to the 1 mbar pressure level. To determine $R_p$ as a function of $M_p$ and age for these planets, we use an updated version of the model described in \citet{2007ApJ...659.1661F}. (This model provides $R_p$ at a series of grid points in $M_p$, $a$ and age, and we linearly interpolate between model grid points.) Although recent studies of evaporation of exoplanet atmospheres suggest taking $R_p$ at the 1 mbar level is the most appropriate choice \citep{2004Icar..170..167Y}, previous studies suggest $R_p$ should be taken at about the 0.1 nbar level \citep{1981Icar...48..150W}. This choice would significantly increase our calculated mass loss rate because $R_p$ would be much larger than what we've chosen, and our estimates here would provide lower limits on the total mass lost. \\
\indent For $M_p < 0.018$ $M_{Jup}$, the previous model for $R_p$ breaks down, so, for planets with such masses, we determine $R_p$ by assuming the planet's density remains constant at its value when $M_p = 0.018$ $M_{Jup}$. This assumption undoubtedly introduces some error into the value for $R_p$, but the relationship of radius to mass remains uncertain for planets transitioning from gas giants to rocky planets with atmospheres of several Earth masses. The atmospheric structure for these planets is a complicated function of atmospheric composition \citep{2010arXiv1001.0976M, 2009arXiv0912.3288R}. Our approximation gives roughly the correct radii for Neptune and Neptune-like extra-solar planets and has been applied before \citep{2008ApJ...678.1396J}. Thus, it should suffice for the current study. We've also assumed that the mass loss has a negligible effect on the planet's thermal evolution, consistent with previous work (\emph{e.g.}, \citealt{1998ApJ...500..428T}).
\subsection{Orbital Evolution Model}
\indent To model the effects of tides on the orbit, we use a standard tidal model that has been applied in many studies of solar and extra-solar systems \citep{1963MNRAS.126..257G, 1966Icar....5..375G, 1996ApJ...470.1187R, 2008CeMDA.101..171F, 2008ApJ...678.1396J}. Like many other close-in exoplanets, CoRoT-7 b raises significant tidal bulges on its host star. For planets that revolve more quickly on circular orbits than their host stars rotate, like CoRoT-7 b, the tidal bulge on the host star lags behind the planet. This interaction results in a torque that spins up the host star and reduces the orbital semi-major axis $a$ as:
\begin{equation}
\frac{da}{dt} = -\frac{9}{2} (G/M_*)^{1/2} \frac{R_* M_p}{Q_*^{\prime}} a^{-11/2}
\label{eqn:dadt}
\end{equation}
where $M_*$ is the stellar mass, $R_*$ is the stellar radius, and $Q_*^{\prime}$ is the modified tidal dissipation parameter for the star \citep{1966Icar....5..375G, 2008ApJ...678.1396J, 2007ApJ...661.1180O}. Because constraints on $Q_{*}^{\prime}$ are poor (\emph{e.g.}, \citealt{2008ApJ...678.1396J}), we consider a range of values: $10^5$, $10^6$, and $10^7$.\\
\indent In using Equation \ref{eqn:dadt}, we assume CoRoT-7 b's orbital eccentricity has been negligible since $t = 0.1$ Gyr. CoRoT-7 c might play a role in planet b's orbital evolution by pumping up its orbital eccentricity \citep{2010ApJ...709L..95B}. However, tides probably keep CoRoT-7 b's eccentricity damped to small values. We also assume planet c has experienced negligible mass loss and tidal evolution over its lifetime, due to its greater distance from the host star. Orbital evolution at that distance for a planet with a mass as small as CoRoT-7 c's minimum mass is also much smaller \citep{2008ApJ...678.1396J}. \\
\indent To constrain CoRoT-7 b's initial mass ($M_{p, init}$) and semi-major axis ($a_{init}$), we numerically integrate Equations \ref{eqn:dMpdt} and \ref{eqn:dadt} together, backward and forward in time, and continually evaluate all the other equations along the way. We consider the full range of current system parameters allowed by observational uncertainty, and run a unique integration for each combination of system parameters. That is, we run models with CoRoT-7 b's current mass $M_{p, cur} = $ 4.0, 4.8 and 5.6 $M_{Earth}$ and with its current semi-major axis $a_{cur} =$ 0.01691, 0.0172, and 0.01749 AU. CoRoT-7's age is reported to lie between 1.2 Gyr and 2.3 Gyr (\citet{2009A&A...506..303Q}), so we also run models with the planet's current age equal to 1.2 and 2.3 Gyr. Along with the range of model parameters we also consider, altogether we run more than 10,000 unique calculations.
\section{Results}
\indent Our results show that the coupling of mass loss and orbital evolution may have played a significant role in CoRoT-7 b's history: Loss of planetary mass can reduce the rate of orbital evolution because the rate of orbital evolution scales as the mass (Equation \ref{eqn:dadt}). At the same time, even modest orbital evolution can significantly modify the mass loss history of the planet because the mass loss rate depends so sensitively on the planet-star distance (Equations \ref{eqn:dMpdt} and \ref{eqn:Ktide}). In this section, we first consider the orbital and mass evolution if CoRoT-7 b has always been a solid planet. Then we consider the evolution if CoRoT-7 b started out as a gas giant.
\subsection{Mass loss and orbital evolution if CoRoT-7 b has always been a rocky planet}
\begin{figure}
\includegraphics[width=8in,angle=-90]{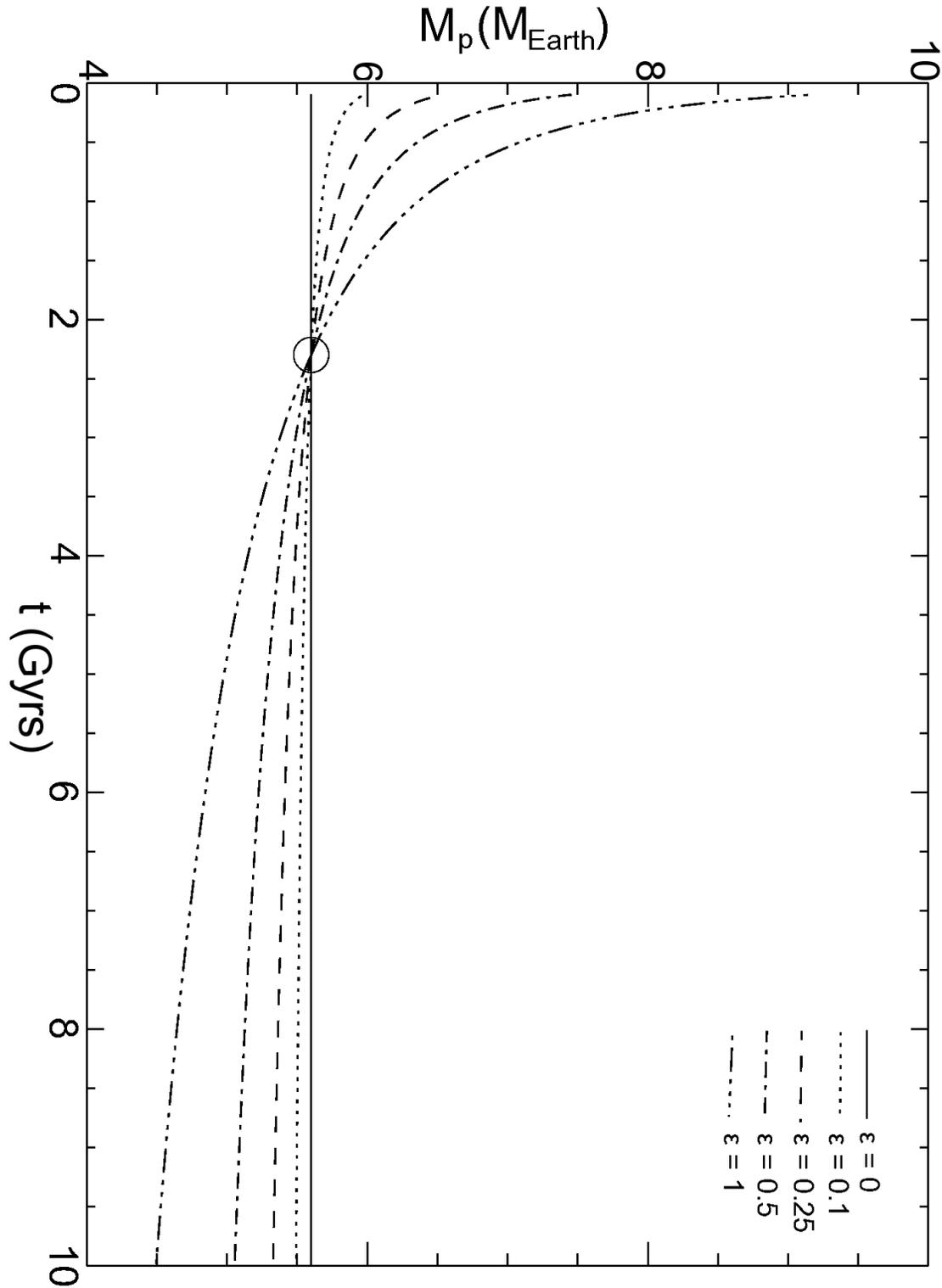}
\caption{Evolution of $M_p$ for CoRoT-7 b, neglecting orbital migration. The empty circle represents an assumed current mass $M_{p, cur} = 5.6$ $M_{Earth}$ and a current age of 2.3 Gyr. The different linestyles correspond to different assumed evaporation efficiencies $\epsilon$, as indicated. For this figure, we've determined $R_p$ as a function of $M_p$ assuming the planet is composed only of rock and iron, with a rock mass fraction (RMF) = 0.719.}
\label{fig:plot_evol_notide_solid}
\end{figure}

\begin{figure}
\includegraphics[height=8in]{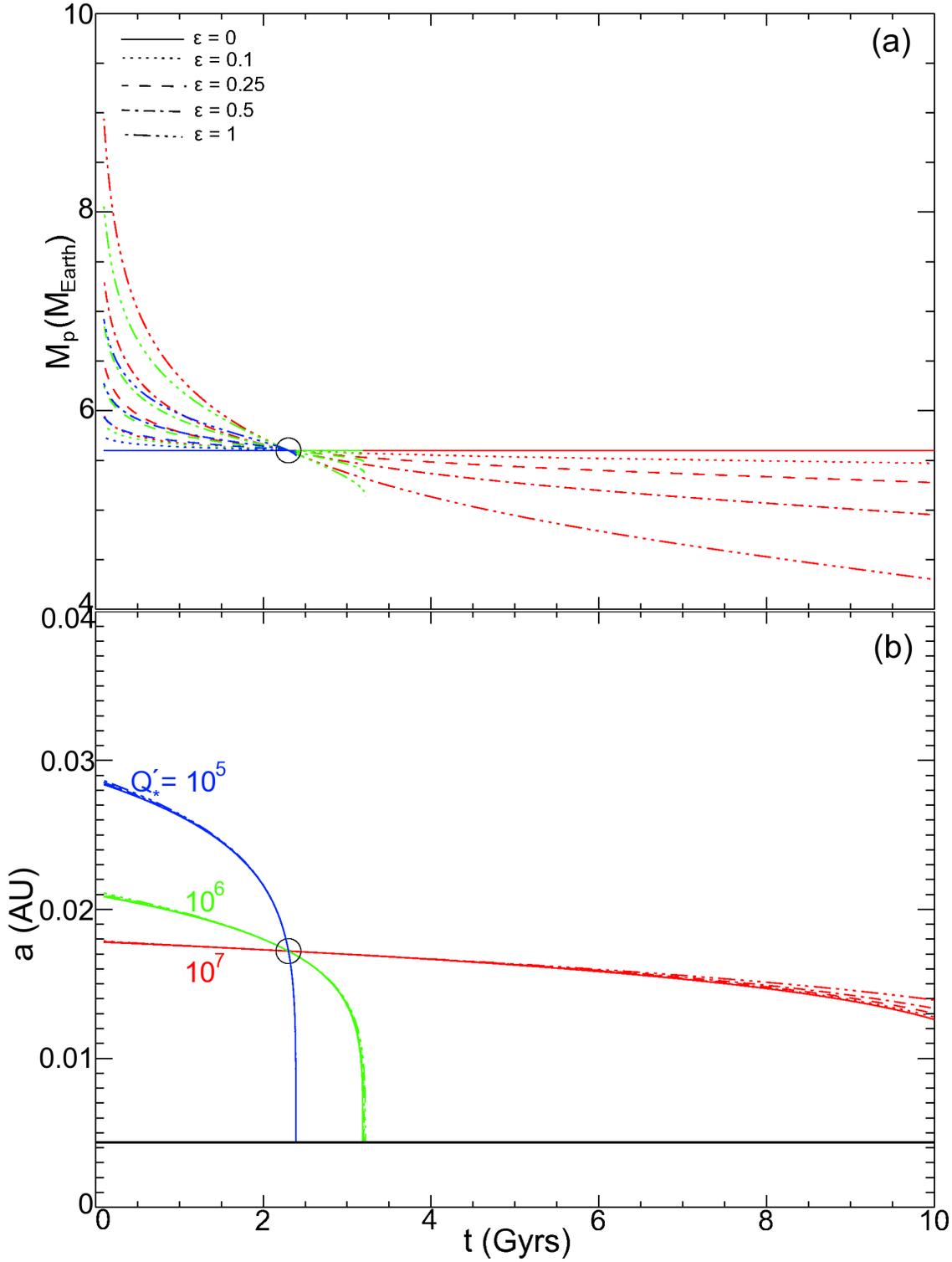}
\caption{Evolution of $M_p$ and $a$ for CoRoT-7 b, including orbital migration. The empty circle in panel (a) represents an assumed current mass $M_{p, cur} = 5.6$ $M_{Earth}$, and in panel (b), an assumed current semi-major axis $a_{cur} = 0.0172$ AU, both for a current age of 2.3 Gyr. Again, we've assumed (RMF) remains constant at 0.719. The different line colors correspond to different assumed values of $Q_*^{\prime}$, as labeled. The solid black line represents the stellar surface.}
\label{fig:plot_evol_tide_solid}
\end{figure}

\begin{figure}
\includegraphics[height=8in]{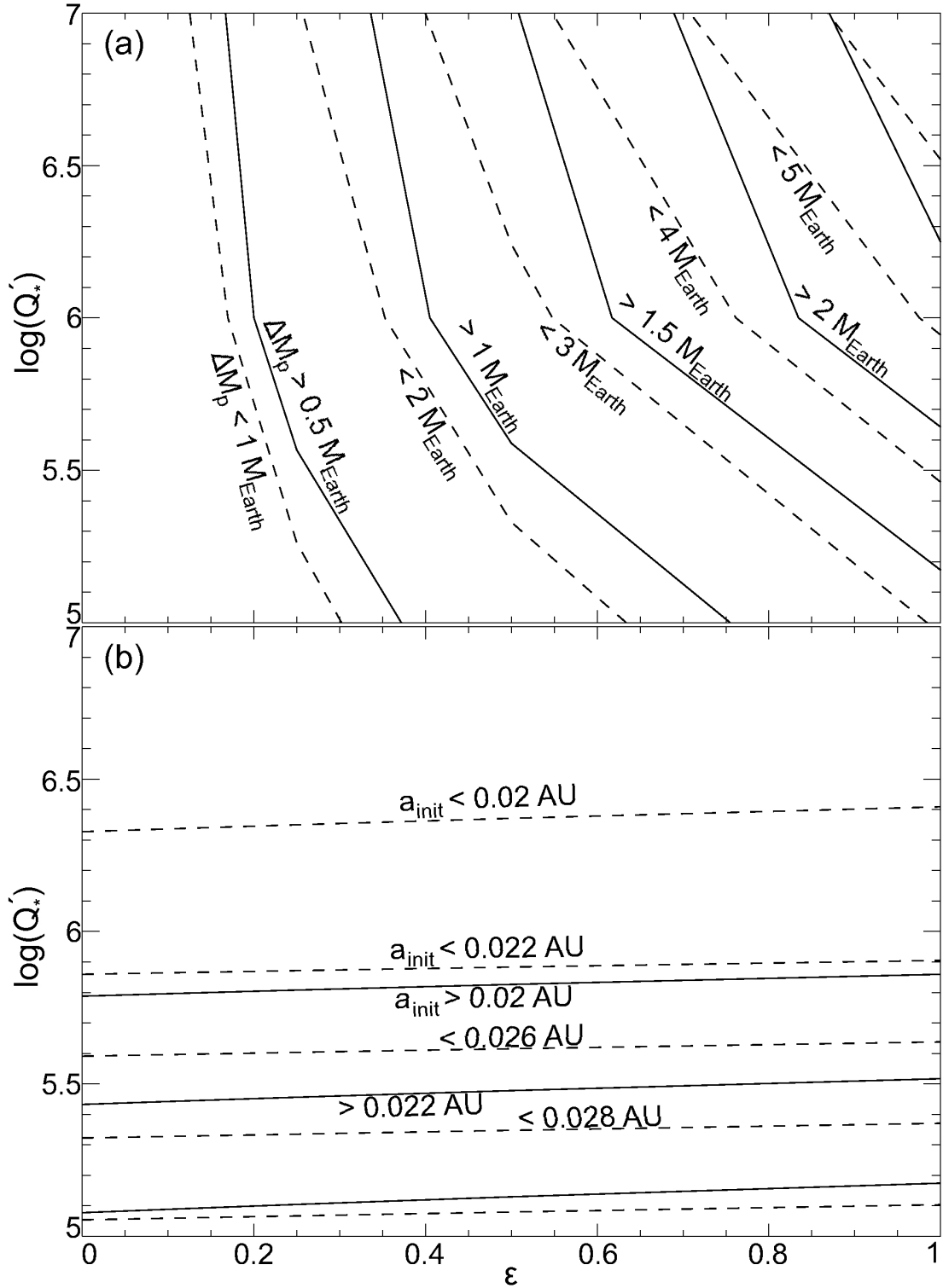}
\caption{Panel (a): Minimum and maximum values for $\Delta M_p = M_{p, init} - M_{p, cur}$ found by considering the range of system and model parameters, as discussed in Section 2. Solid contours show \emph{minimum} values for $\Delta M_p$, and dashed contours show \emph{maximum} values. For example, $\Delta M_p$ lies between 0.5 and 2 $M_{Earth}$ for $\epsilon = 0.4$ and $Q_*^{\prime} = 10^{5.5}$. Panel (b): Minimum and maximum values for $a_{init}$, again by considering a range of parameters. Solid and dashed contours again represent minimum and maximum values, as in panel (a). For example, $a_{init}$ lies between 0.02 and 0.026 AU for $\epsilon = 0.4$ and $Q_*^{\prime} = 10^{5.5}$.}
\label{fig:smallest_largest_solid}
\end{figure}

\indent First we consider mass loss with negligible orbital evolution. Figure \ref{fig:plot_evol_notide_solid} shows the mass evolution of CoRoT-7 b backward and forward in time, assuming $a_{cur} = 0.0172$ AU, $M_{p, cur} = 5.6$ $M_{Earth}$, and $R_p = 1.6$ $R_{Earth}$, which corresponds to a constant rock mass fraction (RMF) = 0.719. At $t = 0.1$ Gyr, we see that CoRoT-7 b may have started with considerably more than its current mass. For example, if $\epsilon = 1$, $M_{p, init}$ could have been as large as about 9.2 $M_{Earth}$. Even for $\epsilon$ as small as 0.1, $M_{p, init}$ was about 0.5 $M_{Earth}$ larger than $M_{p, cur}$. As time moves forward, $dM_p/dt$ drops, largely as a result of the decrease in $F_{xuv}$, and, by assumption, all mass evolution lines converge on $M_{p, cur}$ = 5.6 $M_{Earth}$ at $t = 2.3$ Gyr. Though the current mass loss rate is smaller than in the past, for $\epsilon = 1$, $dM_p/dt \sim 7\times10^{7}$ kg/s (about 0.4 $M_{Earth}$/Gyr). Obviously, smaller $\epsilon$-values give smaller mass loss rates. As time continues into the future, the mass loss rate continues to drop, and for $\epsilon = 1$, the planet will lose about 1 $M_{Earth}$ in a few billion years.\\
\indent In such proximity to its host star, tides likely do drive significant orbital migration. Figure \ref{fig:plot_evol_tide_solid} illustrates the mass evolution for the same $a_{cur}$, $M_{p, cur}$, and (RMF) values as in the previous figure, but this time the calculation includes orbital evolution for a range of stellar $Q_{*}^{\prime}$. We can see from this figure that orbital migration plays an important role in both the planet's past and future.\\
\indent Starting again at $t =$ 0.1 Gyr and moving forward in time, consider first the red lines, for which $Q_{*}^{\prime} = 10^7$. In this case, CoRoT-7 b's orbit changes little, and $a_{init}$ is only about 0.001 AU larger than $a_{cur}$. Thus, the planet receives more XUV radiation at all points in time than if $Q_{*}^{\prime}$ were smaller. As a consequence, $M_{p, init}$ is larger for larger $Q_{*}^{\prime}$. For example, for $\epsilon = 1$ and $Q_{*}^{\prime} = 10^7$, the planet loses about 4 $M_{Earth}$ between $t =$ 0.1 and 2.3 Gyr. By comparison, for $\epsilon = 1$ and $Q_{*}^{\prime} = 10^6$, the total mass lost in the same period is about 3 $M_{Earth}$. Thus, the total mass lost by CoRoT-7 b depends on $Q_{*}^{\prime}$ and $\epsilon$. \\
\indent Now consider orbital evolution proceeding forward in time starting at $t =$ 2.3 Gyr. Again, by assumption, the mass and orbital evolution lines pass through $M_{p, cur}$ and $a_{cur}$ at $t =$ 2.3 Gyr. CoRoT-7 b's orbit quickly decays, and the planet encounters the stellar surface and is destroyed in less than 2 billion years for $Q_{*}^{\prime} \le 10^6$. In this case, the planet only loses about 0.5 $M_{Earth}$ before it is destroyed. For $Q_{*}^{\prime} = 10^7$, orbital evolution is much slower, and the planet loses more mass: for $\epsilon = 1$, about 1.25 $M_{Earth}$ over the next few billion years. Comparing the $\epsilon = 1$ lines in Figures \ref{fig:plot_evol_notide_solid} and \ref{fig:plot_evol_tide_solid}, we see that the planet loses about 0.25 $M_{Earth}$ more by $t = 10$ Gyr with orbital evolution than without. Also, we see that different choices of $\epsilon$-values result in slightly different orbital evolution rates, as indicated by the divergence of the different linestyles in panel (b) of Figure \ref{fig:plot_evol_tide_solid}.\\
\indent Given the short time CoRoT-7 b has left for $Q_{*}^{\prime} \le 10^6$, we might conclude that $Q_{*}^{\prime}$ is much larger. Otherwise, it might seem unlikely that we would see CoRoT-7 b at all. \citet{2010ApJ...708..224H} propose a similar argument for the planet WASP-19 b, which also seems on the verge of plummeting into its host star for $Q_{*}^{\prime} \le 10^9$. On the other hand, close-in exoplanets may well be commonly destroyed through orbital decay \citep{2009ApJ...698.1357J}, so CoRoT-7 b and WASP-19 b may just be next in a long line of planets to be destroyed. \\
\indent Although the amount of mass lost by CoRoT-7 b depends on $Q_{*}^{\prime}$ and $\epsilon$, the planet's current orbit, mass and radius, as well as its composition, are also important for determining the total mass lost by the planet over its lifetime. However, by considering the full range allowed for these parameters, we can constrain the change in mass and the accompanying change in semi-major axis.\\ 
\indent Figure \ref{fig:smallest_largest_solid} shows the range allowed for the mass lost by CoRoT-7 b ($\Delta M_p = M_{p, init} - M_{p, cur}$) and the range allowed for $a_{init}$, as functions of $Q_{*}^{\prime}$ and $\epsilon$. In panel (a), the solid contours show the \emph{minimum} value for $\Delta M_p$ as a function of $Q_{*}^{\prime}$ and $\epsilon$, while the dashed contours show the \emph{maximum} value. For example, consider $Q_{*}^{\prime} = 10^{5.5}$ and $\epsilon = 0.4$, values suggested by \citet{2008ApJ...678.1396J} and the mass loss rate for HD 209458 b (Section 1), respectively. That point lies to the left of the ``2 $M_{Earth}$'' dashed contour, which indicates the total mass loss cannot exceed 2 $M_{Earth}$. The point also lies to the right of the ``0.5 $M_{Earth}$'' solid contour, which indicates the total mass loss must exceed 0.5 $M_{Earth}$. For the same  $Q_{*}^{\prime}$ and $\epsilon$-values, panel (b) tells us CoRoT-7 b's $a_{init}$ lay between 0.022 and 0.024 AU, at least 30 \% larger than $a_{cur}$. \\
\indent In interpreting Figure \ref{fig:smallest_largest_solid}, panel (a), it's important to keep in mind that a dashed contour provides constraints on $\Delta M_p$ only for the region to the left of the contour, while a solid contour provides constraints only for the region to the right of the contour. For example, consider the region sandwiched between the dashed ``1 $M_{Earth}$'' and the solid ``0.5 $M_{Earth}$'' contours. In this region, we can only conclude the change in mass in this region is less than 2 $M_{Earth}$ because it lies to the left of the dashed ``2 $M_{Earth}$'' contour. The contours nearest this region do not directly provide constraints on this region. Otherwise, we might conclude paradoxically that, in this region, the total mass lost is greater than 1 $M_{Earth}$ and less than 0.5 $M_{Earth}$. Similar considerations apply to panel (b), except that constraints from a dashed contour apply to the region \emph{below} the contour, and constraints from a solid contour to the region \emph{above}.\\
\indent Figure \ref{fig:smallest_largest_solid} also illustrates how the total mass lost depends on the orbital evolution and how, to a lesser extent, the orbital evolution depends on the mass loss. In panel (a), for smaller $Q_{*}^{\prime}$, the contour lines become more horizontal. This trend reflects the fact that, for faster orbital evolution, the total mass lost by CoRoT-7 b depends more sensitively on the chosen $Q_{*}^{\prime}$-value, as well as on $\epsilon$. A similar but less pronounced trend is evident in panel (b). The slight upward slope of the lines reflects the fact that, for a fixed $Q_{*}^{\prime}$-value, larger $\epsilon$-values result in more orbital migration during the planet's history. \\
\indent The presence of CoRoT-7 c in a nearby orbit (with $a = 0.046$ AU [\citealt{2009A&A...506..303Q}]) would seem to provide some dynamical constraints on the allowed $M_{p, init}$ and $a_{init}$-values for planet b. We expect that, whatever the original mass and orbit of CoRoT-7 b, gravitational interactions between the planets should never have ejected either planet from the system or changed the ordering of their orbits, \emph{i.e.}, the system has always been dynamically stable. Unfortunately no simple criteria have been found that determines whether a given system is dynamically stable in this way. However, based on \citet{1993Icar..106..247G}, \citet{2006ApJ...647L.163B} showed that the boundary for this kind of stability for a two planet system can be approximated by the following relationship:
\begin{equation}
\alpha^{-3}\left(\mu_1 + \frac{\mu_2}{\delta^2}\right)(\mu_1 + \mu_1\delta)^2 > 1 + 3^{4/3}\frac{\mu_1\mu_2}{\alpha^{4/3}}
\label{eqn:stability}
\end{equation}
where $\mu_i$ is the ratio of planet $i$'s mass to the total system mass, $\alpha = \mu_1 + \mu_2$, and $\delta = (a_2/a_1)^2$. (We've assumed the orbits are circular.) Index 1 refers to the inner planet (b), 2 to the outer planet (c). For planet masses and semi-major axes that satisfy the inequality, the corresponding system is likely to be dynamically stable. For planet masses and semi-major axes that do not satisfy the inequality, the stability is undetermined. Inequality (\ref{eqn:stability}) also does not apply to two planet systems in a low-order mean motion resonance.\\
\indent All $M_{p, init}$ and $a_{init}$-values represented in Figure \ref{fig:smallest_largest_solid} satisfy the inequality, which suggests the planetary system would have been stable for all cases indicated in the figure. However, our calculations do suggest that CoRoT-7 b may have passed through a 3:1 mean motion resonance, as its $a$-value passed through 0.022 AU. Inequality (\ref{eqn:stability}) does not provide any information about the stability of such a resonance encounter, so N-body modeling that includes tidal damping would be required to determine the stability. However, previous studies of mean motion resonances in satellite systems (\emph{e.g.}, \citep{2007DDA....38.0304H}) suggest that, if CoRoT-7 b encountered such a resonance, the planet could migrate through it without being trapped (although the eccentricity could be temporarily boosted). Future studies should address such scenarios.
\subsection{Mass loss and orbital evolution if CoRoT-7 b is the remnant of a gas giant}
\begin{figure}
\includegraphics[width=8in,angle=-90]{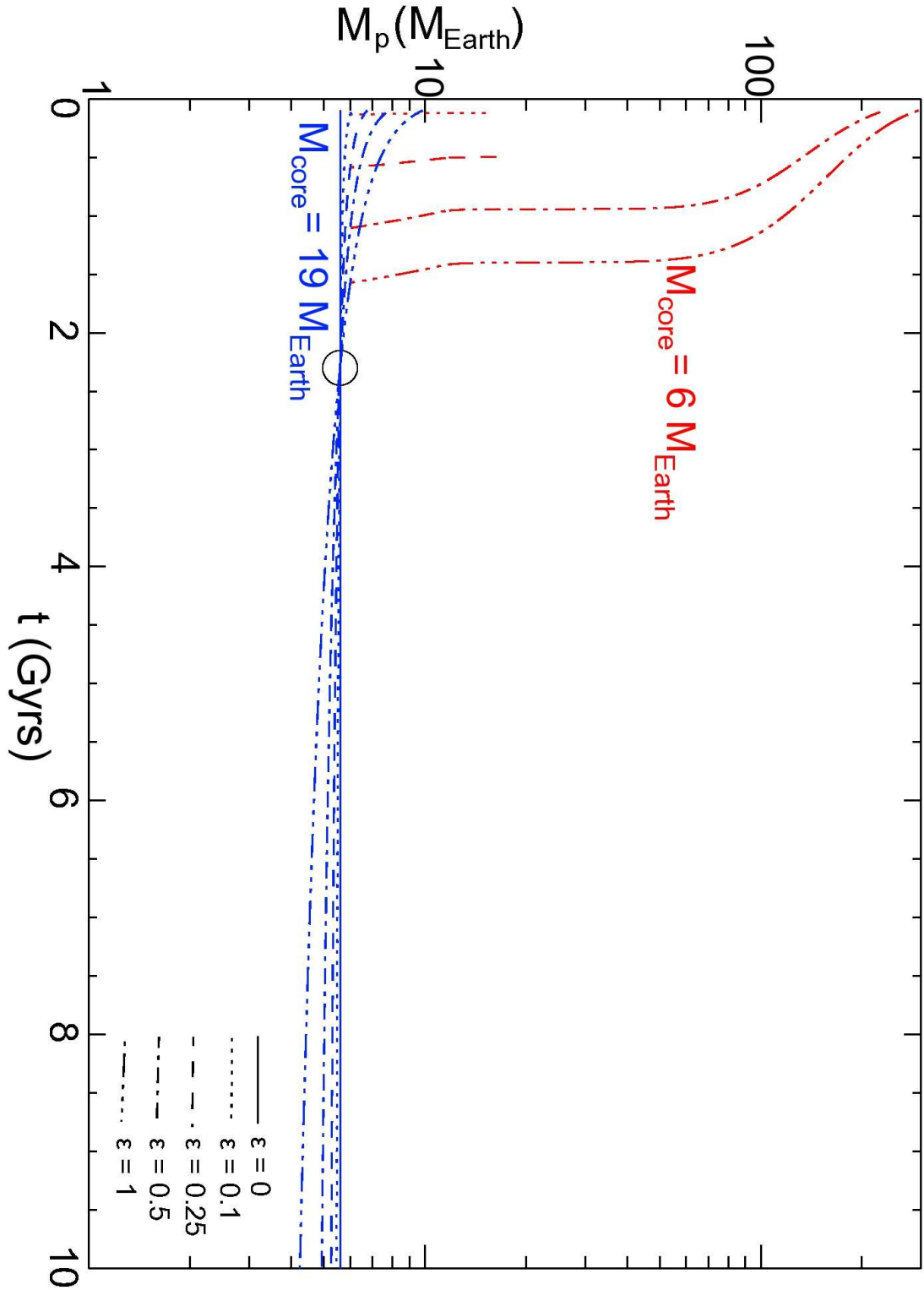}
\caption{Evolution of $M_p$, neglecting orbital migration, for different assumed core masses $M_{core}$, as indicated by the different line colors. The empty circle has the same meaning as in previous figures. While $M_p < M_{core}$ and the planet is losing mass in the form of rocky material, we assumed a constant (RMF) = 0.719.}
\label{fig:plot_evol_notide_gaseous}
\end{figure}

\begin{figure}
\includegraphics[height=8in]{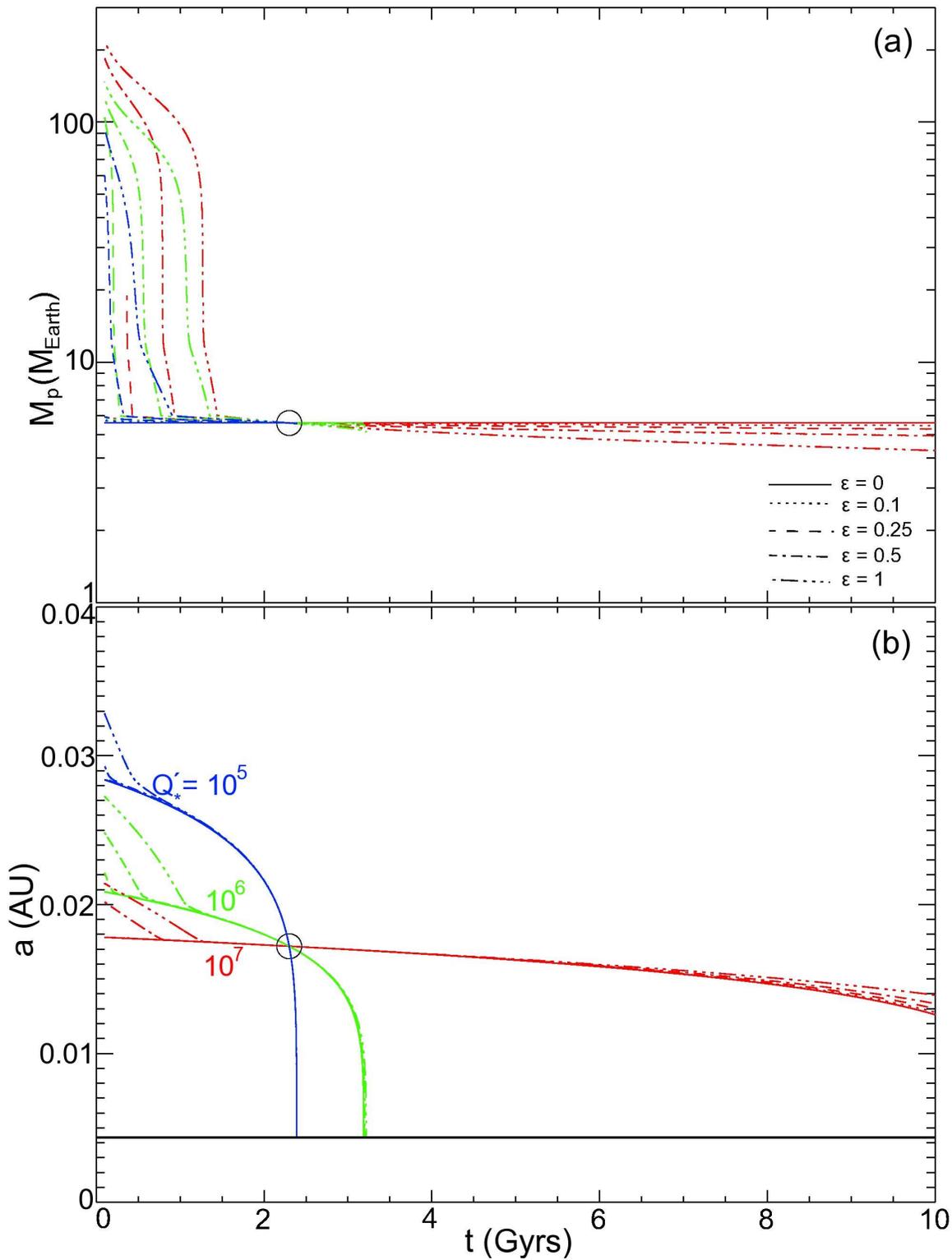}
\caption{Evolution of $M_p$ and $a$ for CoRoT-7 b, including orbital migration. Here, we've assumed (RMF) = 0.719 and $M_{core} = 6$ $M_{Earth}$.}
\label{fig:plot_evol_tide_gaseous}
\end{figure}

\begin{figure}
\includegraphics[height=8in]{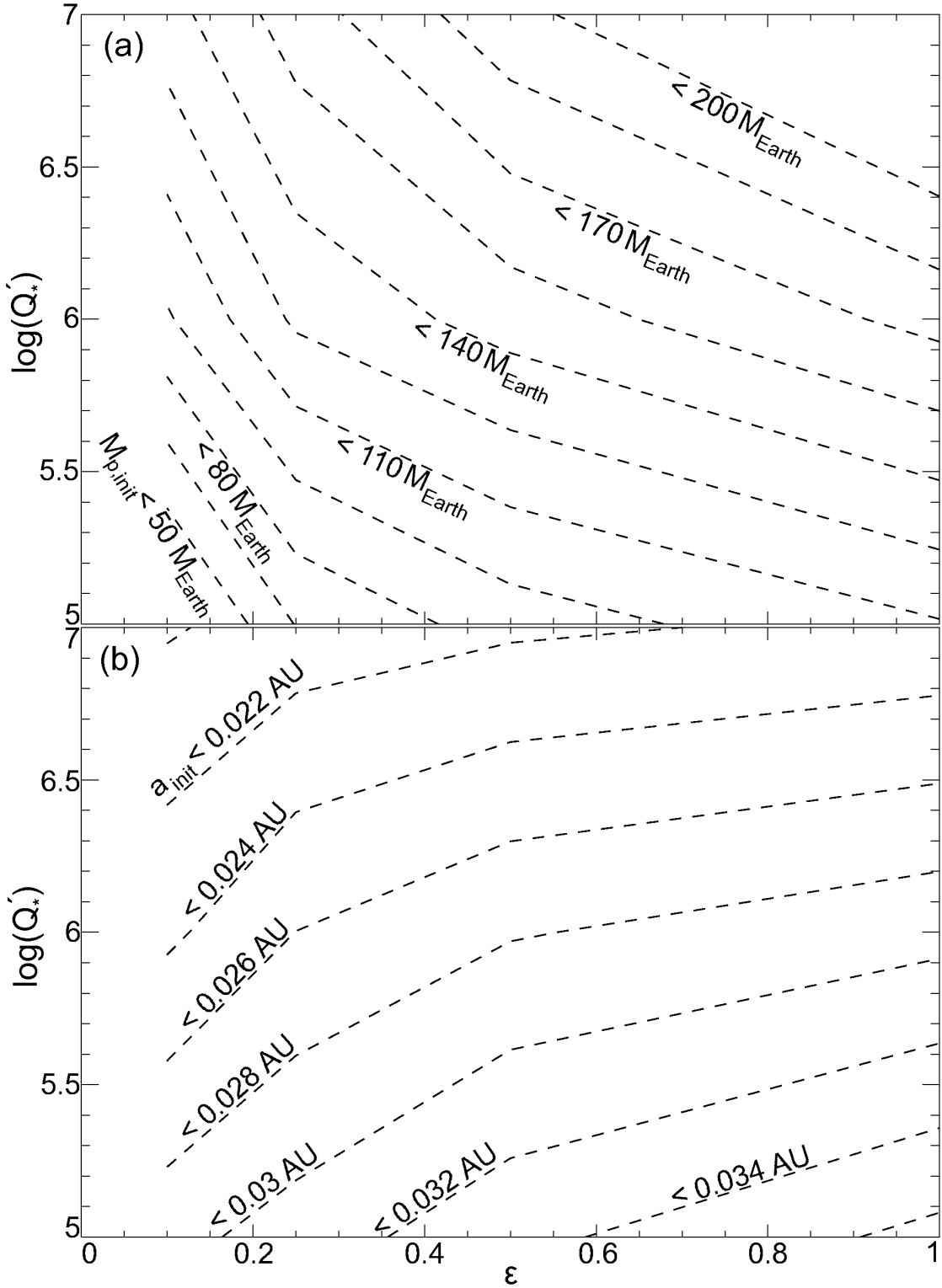}
\caption{The maximum values allowed for CoRoT-7 b's initial mass $M_{p, init}$ and initial semi-major axis $a_{init}$. Dashed contours have meanings similar to the dashed contours in Figure \ref{fig:smallest_largest_solid}. For example, $M_{p, init} < 110 M_{Earth}$ and $a_{init} < 0.03 AU$ for $\epsilon = 0.4$ and $Q_*^{\prime} = 10^{5.5}$, but the contours provide no constraints on the minimum initial values.}
\label{fig:largest_gaseous}
\end{figure}

\indent If CoRoT-7 b began instead as a gas giant planet, the total mass lost may have been much larger. Recall, however, that determining the mass of the original atmosphere requires knowledge of the original core mass. Because, in our model, we allow the solid mass of the planet to evaporate after the original atmosphere evaporates, we can't estimate uniquely both the original core mass and the mass of the atmosphere. To circumvent this problem, we consider a range of values for the original mass of the core $M_{core}$. In the calculations described below, as we run time backward, we allow the solid mass of the planet to increase until it reaches the chosen $M_{core}$. Going back in time past that point, we allow the atmospheric mass to increase until $t = 0.1$ Gyr. In our models, while $M_p \ge M_{core}$, we assume the planet is a gas giant, with an H/He envelope. Once $M_p$ drops below $M_{core}$, we assume the planet has completely shed its H/He envelope and is only a denuded rocky/icy body. The critical core mass above which a rocky/icy body will begin to accrete a significant H/He envelope from the protoplanetary disk may span the range from 1 to more than 10 $M_{Earth}$ (\emph{e.g.}, \citealt{2001ApJ...553..999I}).\\ 
\indent To illustrate the importance of our choice of $M_{core}$, Figure \ref{fig:plot_evol_notide_gaseous} shows mass evolution for two different assumed $M_{core}$-values: 6 (red lines) and 19 $M_{Earth} $ (blue lines). The calculations use the same model parameters as in Figures \ref{fig:plot_evol_notide_solid} and \ref{fig:plot_evol_tide_solid} and neglects orbital evolution. From $t = 2.3$ Gyr forward, the mass evolution for both lines is the same as in previous figures. In the past, though, there are important differences. \\ 
\indent Consider first the blue lines, for which we've set $M_{core} =$ 19 $M_{Earth}$. By assumption, while $M_p < M_{core}$, the planet is only a rocky/icy body with no H/He envelope. For this calculation, we see that the planet's original mass $M_{p, init}$ was always less than $19 M_{Earth}$. Thus, if we require that the planet's initial mass must have been at least $19 M_{Earth}$ to accrete a gaseous envelope from the protoplanetary disk, then this model planet never accreted such an envelope. As a consequence, the planet's radius $R_p$ is always much smaller than it would be if it managed to accrete an atmosphere. By comparison, for the $M_{core} =$ 6 $M_{Earth}$ case (red lines), the solid component of the planet was at least 6 $M_{Earth}$ about two billion years ago for all $\epsilon \ge 0.1$. Thus, according to our assumptions, the planet may have started with a significant gaseous envelope and, consequently, its radius would have been much larger during those times than if it never had an atmosphere, as for the previous model planet. As an example, for $\epsilon = 1$, the atmospheric mass could have been as large as about 200 $M_{Earth}$.\\
\indent For the models we consider in this subsection, our choice of $\epsilon$ plays a role that it did not play in the previous subsection in determining the qualitative nature of past mass loss. As expected, for smaller $\epsilon$-values, the original mass of the planet was smaller. Also, because we assume the planet shed its original atmosphere before $t =$ 2.3 Gyr, the smaller mass loss rates mean that the loss of the original atmosphere must have occurred further in the past. For example, for $\epsilon = 0.5$ and 1.0, the model planet not only started with much larger atmospheric masses than for smaller $\epsilon$, the gaseous envelope was shed as recently as 500 million years ago. By contrast, for $\epsilon = 0.25$ and 0.1, the original atmosphere's mass was less than several $M_{Earth}$ and must have been shed at least 1.5 billion years ago.\\ 
\indent However, we see that the lines for $\epsilon = 0.25$ and 0.1 stop before reaching $t = 0.1$ Gyr. The lines stop because, for slightly more massive atmospheres, the radii of these planets at these early times would have exceeded the Roche lobe, and our mass loss model would break down. By contrast, the model planets with $\epsilon = 0.5$ and 1 are massive enough at these early times that their atmospheres remain bound. This dichotomy is a consequence of the thermal evolution of the planets' interiors. At early times, the interiors of our model gas giants are much warmer than at later times, and the planets experience significant cooling and contraction as they age \citep{2005AREPS..33..493G, 2007ApJ...659.1661F}. Before $t \sim$ 1 Gyr, an atmosphere for a planet of only several $M_{Earth}$ would be so warm and the planet's surface gravity so small that the planet's radius would exceed the Roche limit. A much more massive planet has sufficient gravity that most of its atmosphere remains bound.\\
\indent What does all this mean for the history of CoRoT-7 b? For small mass loss efficiencies (small $\epsilon$) and negligible orbital migration, CoRoT-7 b wouldn't have had enough time since its formation to lose more than a few Earth masses of gas through evaporative mass loss alone. Instead, if the planet started as a gas giant with a mass similar to Neptune's and orbital migration has been negligible, CoRoT-7 b probably underwent a period of Roche lobe overflow early in its life. \\
\indent If CoRoT-7 b has undergone significant orbital migration, though, it may have avoided this early phase of Roche lobe overflow because it would have been farther from its host star than it is now. Figure \ref{fig:plot_evol_tide_gaseous} illustrates the planet's mass evolution including orbital migration for an assumed $M_{core} = 6 M_{Earth}$. Going into the future, we see results similar to previous figures. There are again important differences in the past, though.\\ 
\indent As in Figure \ref{fig:plot_evol_tide_solid}, moving forward in time from $t = 0.1$ Gyr, orbital decay increases the insolation received by the planet, promoting mass loss. Consequently, for a fixed $\epsilon$-value, smaller $Q_{*}^{\prime}$-values correspond to smaller $M_{p, init}$ in general. For example, for $Q_{*}^{\prime} = 10^5$ (blue lines) and $\epsilon \le 0.25$, mass loss rates are always small enough that the planet would not have had time to shed a large gaseous envelope by $t =$ 2.3 Gyr.\\
\indent Figure \ref{fig:plot_evol_tide_gaseous} also shows the important feedback on the orbital migration resulting from substantial mass loss. Panel (b) shows that, for the planets that begin with substantial gaseous envelopes, orbital decay abruptly slows once the planets lose their massive atmospheres. These results show that, if CoRoT-7 b has undergone enough orbital migration, it may have avoided an early phase of Roche lobe overflow. The results also suggest that, in order to retain their atmospheres, any very close-in Neptune-like planets found orbiting sun-like stars must have started farther away from their stars. (Such restrictions may not apply to Gl 1214 b because it orbits a much less massive star than CoRoT-7 [\citealt{2009Natur.462..891C}].)\\
\indent By running many such models, we can constrain the original mass and semi-major axis of a CoRoT-7 b that began as a gas giant. Without knowledge of $M_{core}$, though, we can only estimate the \emph{maximum} value for $M_{p, init}$. Also, because Roche lobe overflow occurs for many of our model planets if they undergo little orbital migration, we can only meaningfully estimate the \emph{maximum} value for $a_{init}$ as well. \\
\indent Figure \ref{fig:largest_gaseous} shows the results of such an analysis. (Because here we are only interested in the possibility that CoRoT-7 b was a gas giant, we exclude models with $\epsilon = 0$ for which there is no mass evolution.) In this figure, we see the same trends indicated in previous figures, although they are more pronounced in this figure as a result of the larger masses involved. For the nominal values $Q_{*}^{\prime} = 10^{5.5}$ and $\epsilon = 0.4$, from panel (a), we see CoRoT-7 b could have started with $M_{p, init}$ no larger than $110$ $M_{Earth}$. For these same model parameters, we see in panel (b) $a_{init} \le 0.028$ AU. Thus CoRoT-7 b may have undergone significant orbital migration over its lifetime. Holding $Q_{*}^{\prime}$ fixed at $10^{5.5}$, we see that larger $\epsilon$-values allow larger $M_{p, init}$, but the contours flatten. This flattening indicates that an increased mass evolution rate is accompanied by an increased orbital evolution rate, which helps to slow the mass evolution -- the same feedback seen in Figure \ref{fig:smallest_largest_solid}. Holding $\epsilon$ fixed at 0.4, we see that larger $Q_{*}^{\prime}$-values likewise allow larger and larger $M_{p, init}$.\\
\indent We can again apply our simple stability criterion, Inequality (\ref{eqn:stability}), to determine whether any of the $M_{p, init}$ and $a_{init}$ values in Figure \ref{fig:largest_gaseous} can be ruled out on dynamical grounds. All values shown, however, satisfy the inequality, suggesting we cannot use dynamical considerations to rule out the possibility that CoRoT-7 b was a gas giant. Also, as in Figure \ref{fig:smallest_largest_solid}, passage through mean motion resonances may have occurred.
\section{Discussion}
\indent Our study suggests CoRoT-7 b may have begun with a much larger mass than it has today, perhaps even with a substantial gaseous component. A more precise determination of CoRoT-7 b's composition may allow us to corroborate or refute the suggestion that the planet began as a gas giant. In particular, the large mass loss rates indicated by our calculations suggest CoRoT-7 b may currently have an extended exosphere of vaporized material, observations of which may provide clues to the planet's composition. Our results also suggest the coupling between orbital evolution and mass loss may have shaped the observed population of close-in exoplanets. Although our calculations have elucidated the roles of orbital evolution and mass loss in the origin of CoRoT-7 b, further work will be needed to test our assumptions and consider other potentially important physical processes.\\
\indent For CoRoT-7 b and other close-in low-mass planets, it may be possible to distinguish remnant cores of gas giants based on estimates of their masses and radii \citep{2008MNRAS.384..663R}. Many previous studies have developed constraints on the compositions of exoplanets made of solid materials based on estimates of their masses and radii \citep{2007ApJ...665.1413V, 2007ApJ...659.1661F, 2007Icar..191..337S, 2008ApJ...673.1160A}. These studies show that determining the compositions of solid planets of end-member compositions (\emph{e.g.}, planets made entirely of ice or iron) is relatively easy but determining the compositions of planets with mixed compositions is more difficult. Moreover, it may be hard to conclude that a solid exoplanet is indeed a remnant core because the compositions of the core of a gas giant may vary from planet to planet, in response to the various formation conditions \citep{1996Icar..124...62P}. \\
\indent In any case, if a close-in solid planet were found to have a substantial volatile component, it is unlikely to have formed \emph{in situ} and is probably the remnant of a more massive planet, perhaps a gas giant \citep{2008MNRAS.384..663R}. In fact, such considerations strongly suggest the recently discovered planet GJ 1214 b \citep{2009Natur.462..891C} may be the remnant of a much larger planet. Given the planet's apparent density, recent studies suggest the planet may have an H/He envelope which is only 0.05\% of the planet's mass \citep{2010arXiv1001.0976M} or a water layer that is almost 50\% of the planet's mass \citep{2009arXiv0912.3288R}. Also, \citet{2009Natur.462..891C} calculate that the current mass loss rate for GJ 1214 b may be sufficient to remove its entire gaseous envelope in about 700 Myrs, suggesting the planet probably started with a larger mass. With its very close-in orbit ($a = 0.0179$ AU [\citealt{2009Natur.462..891C}]), the coupling between mass loss and orbital evolution from tidal damping would have played an important role in GJ 1214 b's history. (We do not include GJ 1214 b here, in part, because our model for stellar XUV radiation is not well-suited for M-dwarf stars, like GJ 1214.)\\
\indent Recently \citet{2009arXiv0907.3067V} considered the constraints on CoRoT-7 b provided its mass and radius. They suggested that the available constraints are consistent with a wide range of compositions and internal structures, from a planet made entirely of silicates to one with a massive water vapor atmosphere. They also considered the possibility that CoRoT-7 b is the remnant of an evaporated gas giant and showed that the planet may have lost many Earth masses of gaseous material, consistent with our results. That study did not include tidal migration, which we've shown may be important.\\
\indent Other recent studies cast doubt on the possibility that CoRoT-7 b started out as a Jupiter-like gas giant. \citet{2009ApJ...693...23M} suggested that complete evaporation of a Jupiter-mass planet's atmosphere may require unrealistically large mass loss efficiencies. \citet{2009A&A...506..399L} suggested that, even if CoRoT-7 b began with a significant gaseous envelope, its original mass was probably less than Neptune's for the same efficiencies we've used here. However, we find that CoRoT-7 b may be the remnant core of a gas giant planet with a mass as large as Saturn's even for relatively small mass loss efficiencies. The reasons for this disagreement probably lie in differences in the model details.\\
\indent For example, \citet{2009A&A...506..399L} considered a model very similar to ours (although their model for the stellar flux was slightly different), but did not apply their model to a planet as close to its star as CoRoT-7 b, which receives more insolation and for which the ``Roche lobe effect'' is more important. Thus it may experience more rapid mass loss than the planets modeled by \citet{2009A&A...506..399L}. In any case, it will be interesting to reconsider the results from previous studies by including the effects of orbital evolution as we have done.\\
\indent If vaporization of CoRoT-7 b continues today, CoRoT-7 b may be surrounded by a dense exosphere, similar to Mercury's. \citet{2009ApJ...703L.113S} found thermal vaporization of rocky material could produce a column abundance of sodium of perhaps $10^{19} cm^{-2}$, almost $10^8$ greater than the abundance of sodium for Mercury (\emph{e.g.}, \citealt{2007SSRv..132..433K}). For comparison, the column abundance of the Earth's atmosphere is $\sim 10^{25} cm^{-2}$. Other processes that generate Mercury's exosphere (\emph{e.g.}, photon-stimulated desorption) would probably produce an even larger column abundance. Although \citet{2009ApJ...703L.113S} did not include processes that might quickly remove vaporized materials from the exosphere (such as photoionization and stellar wind entrainment), such large column abundances for sodium and other materials suggest an exosphere around CoRoT-7 b might be observable. Even if such observations are not possible for CoRoT-7 b, they may be possible for other close-in rocky exoplanets soon to be discovered.\\
\indent Our results also suggest very close-in Neptune-like planets orbiting Sun-like stars may only be able to retain their atmospheres if they have undergone significant orbital decay and started out farther from their host stars. As our understanding of tidal dissipation within stars improves (see, \emph{e.g.}, \citealt{2009ApJ...704..930P}), we may find that $Q_*^{\prime}$-values significantly exceed the values considered here. Such a result might suggest we will be unlikely to find very close-in Neptune-like planets because those planets would have rapidly shed their atmospheres through Roche lobe overflow early in their histories.\\
\indent Our results have implications for the population of observed close-in exoplanets as well. For example, \citet{2007ApJ...658L..59H} compared the mass distribution of close-in exoplanets to that of exoplanets far from their host stars and suggested they are statistically identical, and that the mass distribution expected if significant evaporative mass loss from close-in exoplanets \emph{had} occurred is inconsistent with the observed distribution. However, the sample size of that study was limited (the number of close-in exoplanets has more than doubled since that paper was written), and the disagreement between observations and the expectations from mass loss models was marginal (less than twice the observational standard deviation). More recently, \citet{2009MNRAS.396.1012D} noted a lack of planets with small masses and short orbital periods and suggested evaporation has removed these planets. However, both of these previous studies neglected the effects of orbital evolution. \citet{2009ApJ...698.1357J} noted a similar gap in the distribution of planetary semi-major axes and ages and found the gap is consistent with tidal destruction of planets, but that study ignored the evaporative mass loss that would accompany the orbital evolution. \\
\indent Our results here suggest that the most massive close-in exoplanets lose little mass, but will experience the most rapid orbital evolution. Smaller planets will lose more mass, but experience little orbital evolution. Together these processes may cause the most massive planets to quickly crash into their host stars without losing much mass. The least massive planets could be completely evaporated without much orbital evolution. Planets with intermediate masses may experience both significant orbital evolution and mass loss. Such a scenario would undoubtedly influence the mass distribution for close-in exoplanets. However, the exact balance between loss and replenishment could be complicated and merits further study. \\
\indent Our results have shown the importance of considering tides when studying mass loss of close-in planets and particularly the role tidal considerations can play in constraining the original masses of close-in planets. However, a number of issues remain to be addressed. For example, the dependence of $R_p$ on mass for planets with gaseous envelopes of only several Earth masses is uncertain. For these planets, $R_p$ likely depends in a complex way on the envelope's composition, but the composition of such planets is poorly constrained and may span a wide range. In fact, for such planets, of which GJ 1214 b may be an example, knowledge of the planet's density is not sufficient for determining the planet's composition \citep{2009arXiv0912.3288R, 2010arXiv1001.0976M}. Moreover, significant atmospheric mass loss from gas giants likely results in a time-variable atmospheric composition, as the lightest constituents may be lost first. Studying these details may be especially important for understanding the origins of exoplanets with low masses, which will likely be found in close-in orbits.\\
\indent Another potentially important effect we've neglected is the channeling of escaping gas by the stellar gravity. The tidal gravity of the host star may cause mass to flow preferentially through the L1 point \citep{2003ApJ...588..509G, 2009ApJ...693...23M}. This effect would be especially important during Roche lobe overflow, when a substantial portion of the planet's atmosphere may become unbound from the planet. Exchange of momentum between the escaping gas and the remaining planet can counteract the inward pull of tides, even causing the planet to recede from the star \citep{2010ApJ...708.1692C}. This effect may significantly modify the orbital evolution we've considered here and may be responsible, at least in part, for the clustering of the orbital periods of close-in exoplanets around three days \citep{2008PASP..120..531C}.
\section{Conclusions}
\indent The first discovered rocky exoplanet, CoRoT-7 b, could actually be the remnant core of a gas giant. The planet's proximity to its host star and strong stellar insolation suggest that the planet may have lost a significant amount of mass over its lifetime. The proximity also suggests tides have played an important role in the planet's history, causing it to migrate inward. By coupling models of evaporative mass loss and tidal evolution, we find that the interplay between orbital decay and mass loss can lead to a wide variety of evolutionary pathways. Using a model that couples these two processes, we obtain constraints on the masses and orbits of gas giants that may have given rise to the CoRoT-7 b we see today. Our results suggest a complex orbital history for the planets in the CoRoT-7 system, possibly including passage through mean-motion resonances, even if CoRoT-7 b has always been a rocky planet.\\
\indent Our calculations have general implications for the histories of both close-in rocky and gaseous planets. Previous studies have pointed out the signatures of evaporation and tidal evolution on the distribution of masses and semi-major axes for gaseous close-in exoplanets, but these studies have not considered the coupling of the two processes. Although our calculations allow CoRoT-7 b to be the remnant rocky core of a gas planet, confirming this hypothesis will require additional evidence. For example, improved constraints on the compositions of gas giant cores might allow us to distinguish remnant rocky cores on the basis of their physical properties. More detailed dynamical studies of CoRoT-7 and other similar systems will provide important constraints on their origins and histories. Whatever the details, mass loss and tidal evolution both probably played key roles in the histories and origins of many close-in exoplanets, rocky or gaseous.\\

\section*{Acknowledgments}
We thank Doug Hamilton, Dan Fabrycky, Maki Hattori, Diana Valencia, Marc Kuchner, and John Debes for useful conversations and suggestions. We also thank our referee for a helpful and thorough review. BJ acknowledges support from the NASA Postdoctoral Program. RB acknowledges funding from NASA Astrobiology Institute's Virtual Planetary Laboratory lead team, supported by NASA under Cooperative Agreement No. NNH05ZDA001C. RG acknowledges support from NASA's Planetary Geology and Geophysics program, grant No. NNG05GH65G.

%
%
\bibliographystyle{mn2e}
\bibliography{mn-jour,bib}

\begin{thebibliography}{}

\bibitem[\protect\citeauthoryear{{Abe} \& {Matsui}}{{Abe} \&
  {Matsui}}{1986}]{1986JGR....91..291A}
{Abe} Y.,  {Matsui} T.,  1986, J. Geophys. Res., 91, 291

\bibitem[\protect\citeauthoryear{{Adams}, {Seager} \& {Elkins-Tanton}}{{Adams}
  et~al.}{2008}]{2008ApJ...673.1160A}
{Adams} E.~R.,  {Seager} S.,    {Elkins-Tanton} L.,  2008, Astrophys. J., 673,
  1160

\bibitem[\protect\citeauthoryear{{Baglin}, {Auvergne}, {Boisnard}, {Lam-Trong},
  {Barge}, {Catala}, {Deleuil}, {Michel} \& {Weiss}}{{Baglin}
  et~al.}{2006}]{2006cosp...36.3749B}
{Baglin} A.,  {Auvergne} M.,  {Boisnard} L.,  {Lam-Trong} T.,  {Barge} P.,
  {Catala} C.,  {Deleuil} M.,  {Michel} E.,    {Weiss} W.,  2006, in 36th
  COSPAR Scientific Assembly Vol.~36 of COSPAR, Plenary Meeting, {CoRoT: a high
  precision photometer for stellar ecolution and exoplanet finding}.
pp 3749--+

\bibitem[\protect\citeauthoryear{{Baraffe}, {Chabrier}, {Barman}, {Selsis},
  {Allard} \& {Hauschildt}}{{Baraffe} et~al.}{2005}]{2005A&A...436L..47B}
{Baraffe} I.,  {Chabrier} G.,  {Barman} T.~S.,  {Selsis} F.,  {Allard} F.,
  {Hauschildt} P.~H.,  2005, Astronomy \& Astrophysics, 436, L47

\bibitem[\protect\citeauthoryear{{Baraffe}, {Selsis}, {Chabrier}, {Barman},
  {Allard}, {Hauschildt} \& {Lammer}}{{Baraffe}
  et~al.}{2004}]{2004A&A...419L..13B}
{Baraffe} I.,  {Selsis} F.,  {Chabrier} G.,  {Barman} T.~S.,  {Allard} F.,
  {Hauschildt} P.~H.,    {Lammer} H.,  2004, Astronomy \& Astrophysics, 419,
  L13

\bibitem[\protect\citeauthoryear{{Barker} \& {Ogilvie}}{{Barker} \&
  {Ogilvie}}{2009}]{2009MNRAS.395.2268B}
{Barker} A.~J.,  {Ogilvie} G.~I.,  2009, Monthly Notices of the Royal
  Astronomical Society, 395, 2268

\bibitem[\protect\citeauthoryear{{Barnes} \& {Greenberg}}{{Barnes} \&
  {Greenberg}}{2006}]{2006ApJ...647L.163B}
{Barnes} R.,  {Greenberg} R.,  2006, Astrophys. J. Lett., 647, L163

\bibitem[\protect\citeauthoryear{{Barnes}, {Raymond}, {Greenberg}, {Jackson} \&
  {Kaib}}{{Barnes} et~al.}{2010}]{2010ApJ...709L..95B}
{Barnes} R.,  {Raymond} S.~N.,  {Greenberg} R.,  {Jackson} B.,    {Kaib} N.~A.,
   2010, Astrophys. J. Lett., 709, L95

\bibitem[\protect\citeauthoryear{{Ben-Jaffel}}{{Ben-Jaffel}}{2007}]{2007ApJ...%
671L..61B}
{Ben-Jaffel} L.,  2007, Astrophys. J. Lett., 671, L61

\bibitem[\protect\citeauthoryear{{Borucki}, {Koch}, {Basri}, {Batalha},
  {Brown}, {Caldwell}, {Christensen-Dalsgaard}, {Cochran}, {Dunham}, {Gautier},
  {Geary}, {Gilliland}, {Jenkins}, {Kondo}, {Latham}, {Lissauer} \&
  {Monet}}{{Borucki} et~al.}{2008}]{2008IAUS..249...17B}
{Borucki} W.,  {Koch} D.,  {Basri} G.,  {Batalha} N.,  {Brown} T.,  {Caldwell}
  D.,  {Christensen-Dalsgaard} J.,  {Cochran} W.,  {Dunham} E.,  {Gautier}
  T.~N.,  {Geary} J.,  {Gilliland} R.,  {Jenkins} J.,  {Kondo} Y.,  {Latham}
  D.,  {Lissauer} J.~J.,    {Monet} D.,  2008, in {Y.-S.~Sun, S.~Ferraz-Mello,
  \& J.-L.~Zhou} ed., IAU Symposium Vol.~249 of IAU Symposium, {Finding
  Earth-size planets in the habitable zone: the Kepler Mission}.
pp 17--24

\bibitem[\protect\citeauthoryear{{Chang}, {Gu} \& {Bodenheimer}}{{Chang}
  et~al.}{2010}]{2010ApJ...708.1692C}
{Chang} S.,  {Gu} P.,    {Bodenheimer} P.~H.,  2010, Astrophys. J., 708, 1692

\bibitem[\protect\citeauthoryear{{Charbonneau}, {Berta}, {Irwin}, {Burke},
  {Nutzman}, {Buchhave}, {Lovis}, {Bonfils}, {Latham}, {Udry}, {Murray-Clay},
  {Holman}, {Falco}, {Winn}, {Queloz}, {Pepe}, {Mayor}, {Delfosse} \&
  {Forveille}}{{Charbonneau} et~al.}{2009}]{2009Natur.462..891C}
{Charbonneau} D.,  {Berta} Z.~K.,  {Irwin} J.,  {Burke} C.~J.,  {Nutzman} P.,
  {Buchhave} L.~A.,  {Lovis} C.,  {Bonfils} X.,  {Latham} D.~W.,  {Udry} S.,
  {Murray-Clay} R.~A.,  {Holman} M.~J.,  {Falco} E.~E.,  {Winn} J.~N.,
  {Queloz} D.,  {Pepe} F.,  {Mayor} M.,  {Delfosse} X.,    {Forveille} T.,
  2009, Nature, 462, 891

\bibitem[\protect\citeauthoryear{{Chassefi{\`e}re}}{{Chassefi{\`e}re}}{1996}]{%
1996JGR...10126039C}
{Chassefi{\`e}re} E.,  1996, J. Geophys. Res., 101, 26039

\bibitem[\protect\citeauthoryear{{Cumming}, {Butler}, {Marcy}, {Vogt}, {Wright}
  \& {Fischer}}{{Cumming} et~al.}{2008}]{2008PASP..120..531C}
{Cumming} A.,  {Butler} R.~P.,  {Marcy} G.~W.,  {Vogt} S.~S.,  {Wright} J.~T.,
    {Fischer} D.~A.,  2008, Publications of the Astronomical Society of the
  Pacific, 120, 531

\bibitem[\protect\citeauthoryear{{Davis} \& {Wheatley}}{{Davis} \&
  {Wheatley}}{2009}]{2009MNRAS.396.1012D}
{Davis} T.~A.,  {Wheatley} P.~J.,  2009, Monthly Notices of the Royal
  Astronomical Society, 396, 1012

\bibitem[\protect\citeauthoryear{{Deming}, {Seager}, {Winn}, {Miller-Ricci},
  {Clampin}, {Lindler}, {Greene}, {Charbonneau}, {Laughlin}, {Ricker}, {Latham}
  \& {Ennico}}{{Deming} et~al.}{2009}]{2009PASP..121..952D}
{Deming} D.,  {Seager} S.,  {Winn} J.,  {Miller-Ricci} E.,  {Clampin} M.,
  {Lindler} D.,  {Greene} T.,  {Charbonneau} D.,  {Laughlin} G.,  {Ricker} G.,
  {Latham} D.,    {Ennico} K.,  2009, Publications of the Astronomical Society
  of the Pacific, 121, 952

\bibitem[\protect\citeauthoryear{{Erkaev}, {Kulikov}, {Lammer}, {Selsis},
  {Langmayr}, {Jaritz} \& {Biernat}}{{Erkaev}
  et~al.}{2007}]{2007A&A...472..329E}
{Erkaev} N.~V.,  {Kulikov} Y.~N.,  {Lammer} H.,  {Selsis} F.,  {Langmayr} D.,
  {Jaritz} G.~F.,    {Biernat} H.~K.,  2007, Astronomy \& Astrophysics, 472,
  329

\bibitem[\protect\citeauthoryear{{Ferraz-Mello}, {Rodr{\'{\i}}guez} \&
  {Hussmann}}{{Ferraz-Mello} et~al.}{2008}]{2008CeMDA.101..171F}
{Ferraz-Mello} S.,  {Rodr{\'{\i}}guez} A.,    {Hussmann} H.,  2008, Celestial
  Mechanics and Dynamical Astronomy, 101, 171

\bibitem[\protect\citeauthoryear{{Ford} \& {Rasio}}{{Ford} \&
  {Rasio}}{2006}]{2006ApJ...638L..45F}
{Ford} E.~B.,  {Rasio} F.~A.,  2006, Astrophys. J. Lett., 638, L45

\bibitem[\protect\citeauthoryear{{Fortney}, {Marley} \& {Barnes}}{{Fortney}
  et~al.}{2007a}]{2007ApJ...668.1267F}
{Fortney} J.~J.,  {Marley} M.~S.,    {Barnes} J.~W.,  2007a, Astrophys. J.,
  668, 1267

\bibitem[\protect\citeauthoryear{{Fortney}, {Marley} \& {Barnes}}{{Fortney}
  et~al.}{2007b}]{2007ApJ...659.1661F}
{Fortney} J.~J.,  {Marley} M.~S.,    {Barnes} J.~W.,  2007b, Astrophys. J.,
  659, 1661

\bibitem[\protect\citeauthoryear{{Gaidos}, {Haghighipour}, {Agol}, {Latham},
  {Raymond} \& {Rayner}}{{Gaidos} et~al.}{2007}]{2007Sci...318..210G}
{Gaidos} E.,  {Haghighipour} N.,  {Agol} E.,  {Latham} D.,  {Raymond} S.,
  {Rayner} J.,  2007, Science, 318, 210

\bibitem[\protect\citeauthoryear{{Gladman}}{{Gladman}}{1993}]{1993Icar..106..2%
47G}
{Gladman} B.,  1993, Icarus, 106, 247

\bibitem[\protect\citeauthoryear{{Goldreich}}{{Goldreich}}{1963}]{1963MNRAS.12%
6..257G}
{Goldreich} P.,  1963, Monthly Notices of the Royal Astronomical Society, 126,
  257

\bibitem[\protect\citeauthoryear{{Goldreich} \& {Soter}}{{Goldreich} \&
  {Soter}}{1966}]{1966Icar....5..375G}
{Goldreich} P.,  {Soter} S.,  1966, Icarus, 5, 375

\bibitem[\protect\citeauthoryear{{Gu}, {Lin} \& {Bodenheimer}}{{Gu}
  et~al.}{2003}]{2003ApJ...588..509G}
{Gu} P.,  {Lin} D.~N.~C.,    {Bodenheimer} P.~H.,  2003, Astrophys. J., 588,
  509

\bibitem[\protect\citeauthoryear{{Guillot}}{{Guillot}}{2005}]{2005AREPS..33..4%
93G}
{Guillot} T.,  2005, Annual Review of Earth and Planetary Sciences, 33, 493

\bibitem[\protect\citeauthoryear{{Hamilton} \& {Zhang}}{{Hamilton} \&
  {Zhang}}{2007}]{2007DDA....38.0304H}
{Hamilton} D.~P.,  {Zhang} K.,  2007, in AAS/Division of Dynamical Astronomy
  Meeting Vol.~38 of AAS/Division of Dynamical Astronomy Meeting, {An Intuitive
  Explanation of Resonance Dynamics}.
pp 03.04--+

\bibitem[\protect\citeauthoryear{{Hattori}}{{Hattori}}{2008}]{2008PhDT........%
15H}
{Hattori} M.~F.,  2008, PhD thesis, The University of Arizona

\bibitem[\protect\citeauthoryear{{Hebb}, {Collier-Cameron}, {Triaud}, {Lister},
  {Smalley}, {Maxted}, {Hellier}, {Anderson} et~al.,}{{Hebb}
  et~al.}{2010}]{2010ApJ...708..224H}
{Hebb} L.,  {Collier-Cameron} A.,  {Triaud} A.~H.~M.~J.,  {Lister} T.~A.,
  {Smalley} B.,  {Maxted} P.~F.~L.,  {Hellier} C.,  {Anderson} D.~R.,
  et~al., 2010, Astrophys. J., 708, 224

\bibitem[\protect\citeauthoryear{{Holmstr{\"o}m}, {Ekenb{\"a}ck}, {Selsis},
  {Penz}, {Lammer} \& {Wurz}}{{Holmstr{\"o}m}
  et~al.}{2008}]{2008Natur.451..970H}
{Holmstr{\"o}m} M.,  {Ekenb{\"a}ck} A.,  {Selsis} F.,  {Penz} T.,  {Lammer} H.,
     {Wurz} P.,  2008, Nature, 451, 970

\bibitem[\protect\citeauthoryear{{Hubbard}, {Hattori}, {Burrows} \&
  {Hubeny}}{{Hubbard} et~al.}{2007}]{2007ApJ...658L..59H}
{Hubbard} W.~B.,  {Hattori} M.~F.,  {Burrows} A.,    {Hubeny} I.,  2007,
  Astrophys. J. Lett., 658, L59

\bibitem[\protect\citeauthoryear{{Hubbard}, {Hattori}, {Burrows}, {Hubeny} \&
  {Sudarsky}}{{Hubbard} et~al.}{2007}]{2007Icar..187..358H}
{Hubbard} W.~B.,  {Hattori} M.~F.,  {Burrows} A.,  {Hubeny} I.,    {Sudarsky}
  D.,  2007, Icarus, 187, 358

\bibitem[\protect\citeauthoryear{{Ikoma}, {Emori} \& {Nakazawa}}{{Ikoma}
  et~al.}{2001}]{2001ApJ...553..999I}
{Ikoma} M.,  {Emori} H.,    {Nakazawa} K.,  2001, Astrophys. J., 553, 999

\bibitem[\protect\citeauthoryear{{Jackson}, {Barnes} \& {Greenberg}}{{Jackson}
  et~al.}{2009}]{2009ApJ...698.1357J}
{Jackson} B.,  {Barnes} R.,    {Greenberg} R.,  2009, Astrophys. J., 698, 1357

\bibitem[\protect\citeauthoryear{{Jackson}, {Greenberg} \& {Barnes}}{{Jackson}
  et~al.}{2008}]{2008ApJ...678.1396J}
{Jackson} B.,  {Greenberg} R.,    {Barnes} R.,  2008, Astrophys. J., 678, 1396

\bibitem[\protect\citeauthoryear{{Jaritz}, {Endler}, {Langmayr}, {Lammer},
  {Grie{\ss}meier}, {Erkaev} \& {Biernat}}{{Jaritz}
  et~al.}{2005}]{2005A&A...439..771J}
{Jaritz} G.~F.,  {Endler} S.,  {Langmayr} D.,  {Lammer} H.,  {Grie{\ss}meier}
  J.,  {Erkaev} N.~V.,    {Biernat} H.~K.,  2005, Astronomy \& Astrophysics,
  439, 771

\bibitem[\protect\citeauthoryear{{Killen}, {Cremonese}, {Lammer}, {Orsini},
  {Potter}, {Sprague}, {Wurz}, {Khodachenko}, {Lichtenegger}, {Milillo} \&
  {Mura}}{{Killen} et~al.}{2007}]{2007SSRv..132..433K}
{Killen} R.,  {Cremonese} G.,  {Lammer} H.,  {Orsini} S.,  {Potter} A.~E.,
  {Sprague} A.~L.,  {Wurz} P.,  {Khodachenko} M.~L.,  {Lichtenegger} H.~I.~M.,
  {Milillo} A.,    {Mura} A.,  2007, Space Science Reviews, 132, 433

\bibitem[\protect\citeauthoryear{{Lammer}, {Odert}, {Leitzinger},
  {Khodachenko}, {Panchenko}, {Kulikov}, {Zhang}, {Lichtenegger}
  et~al.,}{{Lammer} et~al.}{2009}]{2009A&A...506..399L}
{Lammer} H.,  {Odert} P.,  {Leitzinger} M.,  {Khodachenko} M.~L.,  {Panchenko}
  M.,  {Kulikov} Y.~N.,  {Zhang} T.~L.,  {Lichtenegger} H.~I.~M.,    et~al.,
  2009, Astronomy \& Astrophysics, 506, 399

\bibitem[\protect\citeauthoryear{{Lammer}, {Selsis}, {Ribas}, {Guinan}, {Bauer}
  \& {Weiss}}{{Lammer} et~al.}{2003}]{2003ApJ...598L.121L}
{Lammer} H.,  {Selsis} F.,  {Ribas} I.,  {Guinan} E.~F.,  {Bauer} S.~J.,
  {Weiss} W.~W.,  2003, Astrophys. J. Lett., 598, L121

\bibitem[\protect\citeauthoryear{{Lecavelier des Etangs}, {Vidal-Madjar},
  {McConnell} \& {H{\'e}brard}}{{Lecavelier des Etangs}
  et~al.}{2004}]{2004A&A...418L...1L}
{Lecavelier des Etangs} A.,  {Vidal-Madjar} A.,  {McConnell} J.~C.,
  {H{\'e}brard} G.,  2004, Astronomy \& Astrophysics, 418, L1

\bibitem[\protect\citeauthoryear{{L{\'e}ger}, {Rouan}, {Schneider}, {Barge},
  {Fridlund}, {Samuel}, {Ollivier}, {Guenther} et~al.,}{{L{\'e}ger}
  et~al.}{2009}]{2009A&A...506..287L}
{L{\'e}ger} A.,  {Rouan} D.,  {Schneider} J.,  {Barge} P.,  {Fridlund} M.,
  {Samuel} B.,  {Ollivier} M.,  {Guenther} E.,    et~al., 2009, Astronomy \&
  Astrophysics, 506, 287

\bibitem[\protect\citeauthoryear{{Levrard}, {Winisdoerffer} \&
  {Chabrier}}{{Levrard} et~al.}{2009}]{2009ApJ...692L...9L}
{Levrard} B.,  {Winisdoerffer} C.,    {Chabrier} G.,  2009, Astrophys. J.
  Lett., 692, L9

\bibitem[\protect\citeauthoryear{{Mardling} \& {Lin}}{{Mardling} \&
  {Lin}}{2004}]{2004ApJ...614..955M}
{Mardling} R.~A.,  {Lin} D.~N.~C.,  2004, Astrophys. J., 614, 955

\bibitem[\protect\citeauthoryear{{Miller-Ricci} \& {Fortney}}{{Miller-Ricci} \&
  {Fortney}}{2010}]{2010arXiv1001.0976M}
{Miller-Ricci} E.,  {Fortney} J.~J.,  2010, ArXiv e-prints

\bibitem[\protect\citeauthoryear{{Murray-Clay}, {Chiang} \&
  {Murray}}{{Murray-Clay} et~al.}{2009}]{2009ApJ...693...23M}
{Murray-Clay} R.~A.,  {Chiang} E.~I.,    {Murray} N.,  2009, Astrophys. J.,
  693, 23

\bibitem[\protect\citeauthoryear{{Ogilvie} \& {Lin}}{{Ogilvie} \&
  {Lin}}{2007}]{2007ApJ...661.1180O}
{Ogilvie} G.~I.,  {Lin} D.~N.~C.,  2007, Astrophys. J., 661, 1180

\bibitem[\protect\citeauthoryear{{Penev}, {Sasselov}, {Robinson} \&
  {Demarque}}{{Penev} et~al.}{2009}]{2009ApJ...704..930P}
{Penev} K.,  {Sasselov} D.,  {Robinson} F.,    {Demarque} P.,  2009, Astrophys.
  J., 704, 930

\bibitem[\protect\citeauthoryear{{Pollack}, {Hollenbach}, {Beckwith},
  {Simonelli}, {Roush} \& {Fong}}{{Pollack} et~al.}{1994}]{1994ApJ...421..615P}
{Pollack} J.~B.,  {Hollenbach} D.,  {Beckwith} S.,  {Simonelli} D.~P.,  {Roush}
  T.,    {Fong} W.,  1994, Astrophys. J., 421, 615

\bibitem[\protect\citeauthoryear{{Pollack}, {Hubickyj}, {Bodenheimer},
  {Lissauer}, {Podolak} \& {Greenzweig}}{{Pollack}
  et~al.}{1996}]{1996Icar..124...62P}
{Pollack} J.~B.,  {Hubickyj} O.,  {Bodenheimer} P.,  {Lissauer} J.~J.,
  {Podolak} M.,    {Greenzweig} Y.,  1996, Icarus, 124, 62

\bibitem[\protect\citeauthoryear{{Queloz}, {Bouchy}, {Moutou}, {Hatzes},
  {H{\'e}brard}, {Alonso}, {Auvergne}, {Baglin} et~al.,}{{Queloz}
  et~al.}{2009}]{2009A&A...506..303Q}
{Queloz} D.,  {Bouchy} F.,  {Moutou} C.,  {Hatzes} A.,  {H{\'e}brard} G.,
  {Alonso} R.,  {Auvergne} M.,  {Baglin} A.,    et~al., 2009, Astronomy \&
  Astrophysics, 506, 303

\bibitem[\protect\citeauthoryear{{Rasio}, {Tout}, {Lubow} \& {Livio}}{{Rasio}
  et~al.}{1996}]{1996ApJ...470.1187R}
{Rasio} F.~A.,  {Tout} C.~A.,  {Lubow} S.~H.,    {Livio} M.,  1996, Astrophys.
  J., 470, 1187

\bibitem[\protect\citeauthoryear{{Raymond}, {Barnes} \& {Mandell}}{{Raymond}
  et~al.}{2008}]{2008MNRAS.384..663R}
{Raymond} S.~N.,  {Barnes} R.,    {Mandell} A.~M.,  2008, Monthly Notices of
  the Royal Astronomical Society, 384, 663

\bibitem[\protect\citeauthoryear{{Ribas}, {Guinan}, {G{\"u}del} \&
  {Audard}}{{Ribas} et~al.}{2005}]{2005ApJ...622..680R}
{Ribas} I.,  {Guinan} E.~F.,  {G{\"u}del} M.,    {Audard} M.,  2005, Astrophys.
  J., 622, 680

\bibitem[\protect\citeauthoryear{{Rogers} \& {Seager}}{{Rogers} \&
  {Seager}}{2009}]{2009arXiv0912.3288R}
{Rogers} L.~A.,  {Seager} S.,  2009, ArXiv e-prints

\bibitem[\protect\citeauthoryear{{Schaefer} \& {Fegley}}{{Schaefer} \&
  {Fegley}}{2009}]{2009ApJ...703L.113S}
{Schaefer} L.,  {Fegley} B.,  2009, Astrophys. J. Lett., 703, L113

\bibitem[\protect\citeauthoryear{{Skumanich}}{{Skumanich}}{1972}]{1972ApJ...17%
1..565S}
{Skumanich} A.,  1972, Astrophysical Journal, 171, 565

\bibitem[\protect\citeauthoryear{{Sotin}, {Grasset} \& {Mocquet}}{{Sotin}
  et~al.}{2007}]{2007Icar..191..337S}
{Sotin} C.,  {Grasset} O.,    {Mocquet} A.,  2007, Icarus, 191, 337

\bibitem[\protect\citeauthoryear{{Tian}, {Toon}, {Pavlov} \& {De
  Sterck}}{{Tian} et~al.}{2005}]{2005ApJ...621.1049T}
{Tian} F.,  {Toon} O.~B.,  {Pavlov} A.~A.,    {De Sterck} H.,  2005, Astrophys.
  J., 621, 1049

\bibitem[\protect\citeauthoryear{{Trilling}, {Benz}, {Guillot}, {Lunine},
  {Hubbard} \& {Burrows}}{{Trilling} et~al.}{1998}]{1998ApJ...500..428T}
{Trilling} D.~E.,  {Benz} W.,  {Guillot} T.,  {Lunine} J.~I.,  {Hubbard} W.~B.,
     {Burrows} A.,  1998, Astrophys. J., 500, 428

\bibitem[\protect\citeauthoryear{{Valencia}, {Ikoma}, {Guillot} \&
  {Nettelmann}}{{Valencia} et~al.}{2009}]{2009arXiv0907.3067V}
{Valencia} D.,  {Ikoma} M.,  {Guillot} T.,    {Nettelmann} N.,  2009, ArXiv
  e-prints

\bibitem[\protect\citeauthoryear{{Valencia}, {Sasselov} \&
  {O'Connell}}{{Valencia} et~al.}{2007}]{2007ApJ...665.1413V}
{Valencia} D.,  {Sasselov} D.~D.,    {O'Connell} R.~J.,  2007, Astrophys. J.,
  665, 1413

\bibitem[\protect\citeauthoryear{{Vidal-Madjar}, {D{\'e}sert}, {Lecavelier des
  Etangs}, {H{\'e}brard}, {Ballester}, {Ehrenreich}, {Ferlet}, {McConnell},
  {Mayor} \& {Parkinson}}{{Vidal-Madjar} et~al.}{2004}]{2004ApJ...604L..69V}
{Vidal-Madjar} A.,  {D{\'e}sert} J.,  {Lecavelier des Etangs} A.,
  {H{\'e}brard} G.,  {Ballester} G.~E.,  {Ehrenreich} D.,  {Ferlet} R.,
  {McConnell} J.~C.,  {Mayor} M.,    {Parkinson} C.~D.,  2004, Astrophys. J.
  Lett., 604, L69

\bibitem[\protect\citeauthoryear{{Vidal-Madjar}, {Lecavelier des Etangs},
  {D{\'e}sert}, {Ballester}, {Ferlet}, {H{\'e}brard} \& {Mayor}}{{Vidal-Madjar}
  et~al.}{2003}]{2003Natur.422..143V}
{Vidal-Madjar} A.,  {Lecavelier des Etangs} A.,  {D{\'e}sert} J.,  {Ballester}
  G.~E.,  {Ferlet} R.,  {H{\'e}brard} G.,    {Mayor} M.,  2003, Nature, 422,
  143

\bibitem[\protect\citeauthoryear{{Watson}, {Donahue} \& {Walker}}{{Watson}
  et~al.}{1981}]{1981Icar...48..150W}
{Watson} A.~J.,  {Donahue} T.~M.,    {Walker} J.~C.~G.,  1981, Icarus, 48, 150

\bibitem[\protect\citeauthoryear{{Yelle}, {Lammer} \& {Ip}}{{Yelle}
  et~al.}{2008}]{2008SSRv..139..437Y}
{Yelle} R.,  {Lammer} H.,    {Ip} W.,  2008, Space Science Reviews, 139, 437

\bibitem[\protect\citeauthoryear{{Yelle}}{{Yelle}}{2004}]{2004Icar..170..167Y}
{Yelle} R.~V.,  2004, Icarus, 170, 167

\end{thebibliography}

\onecolumn
\pagebreak

\label{lastpage}
\end{document}